\documentclass[pra,twocolumn,aps,superscriptaddress]{revtex4-1}
\usepackage{bm}%
\expandafter\ifx\csname package@font\endcsname\relax\else
 \expandafter\expandafter
 \expandafter\usepackage
 \expandafter\expandafter
 
 \expandafter{\csname package@font\endcsname}
\fi

\usepackage{array}
\usepackage{amsmath,amssymb}
\usepackage{MnSymbol}
 \usepackage{float}
\usepackage{graphicx}
\usepackage{lipsum}
\usepackage{graphicx}%
\usepackage{subfigure}
\usepackage{mathrsfs}
\usepackage{bm}
\usepackage{mathtools}
\usepackage{epstopdf}
\usepackage{hyperref}

\hypersetup{%
   pdfpagemode=None, 
   pdfstartpage=1,
   pdfmenubar=true,
   pdftoolbar=true,
   colorlinks = true,
   linkcolor=blue,
   citecolor=blue,
   urlcolor=blue,
   bookmarksopen=false
 }

\usepackage[normalem]{ulem}   
\makeindex

\begin{document}
\title{Moir\'e-induced optical non-linearities: Single and multi-photon resonances}
\author{A. Camacho-Guardian}
\affiliation{T.C.M. Group, Cavendish Laboratory, University of Cambridge, JJ Thomson Avenue, Cambridge CB3 0HE, United Kingdom\looseness=-1}
\author{N. R. Cooper}
\affiliation{T.C.M. Group, Cavendish Laboratory, University of Cambridge, JJ Thomson Avenue, Cambridge CB3 0HE, United Kingdom\looseness=-1}
\affiliation{Department of Physics and Astronomy, University of Florence, Via G. Sansone 1, 50019 Sesto Fiorentino, Italy\looseness=-1}
\date{\today}
\begin{abstract} Moir\'e excitons promise a new platform with which to generate and manipulate hybrid quantum phases of light and matter in unprecedented regimes of interaction strength. We explore the properties in this regime, through studies of a  Bose-Hubbard model of excitons coupled to cavity photons. We show that the steady states exhibit a rich phase diagram with pronounced bi-stabilities governed by multi-photon resonances reflecting the strong inter-exciton interactions. 
In the presence of an incoherent pumping of excitons we find that the system can realise one- and multi-photon lasers. 

\end{abstract}
\maketitle
{\it Introduction.-} The ability to tune the optical, electronic, and transport properties in van der Waals heterostructures has opened  the door for the engineering, realisation, and detection of intriguing complex many-body phases of matter~\cite{Manzeli2017}. Much attention has focused on the electronic properties, which show novel quantum phases arising from strong interactions and provide a new testbed for quantum simulation~\cite{Tang2020,Kennes2021}. However, these materials  also hold  interesting opportunities to study new hybrid quantum states of light and matter~\cite{Baek2020,Yu2020}, and offer promising potential optoelectronic applications based on valleytronics~\cite{Liu2016,Schaibley2016,Rasmita2020,Tang2020a,Forg2021} and twistronics~\cite{Hsu2019}. 

In twisted bilayers of transition-metal dichalcogenides (TMD) the emergent moir\'e periodicity~\cite{Jung2014,Zhang2017,Ulstrup2020,Tran2020} has significant effects on the nature of the excitons -- the bound electron-hole pairs which dominate the coupling to light. In general these excitons are hybrids of the intra-layer excitons formed by an electron and hole in the same layer and inter-layer excitons with an electron and a hole belonging to different layers~\cite{Jiang2021,Tartakovskii2020}. Hybrid excitons have attracted much attention as they  possess tuneable features of both intra- and inter-layer character~\cite{Alexeev2019,Ruiz2019,Shimazaki2020}, and have strong mutual repulsion due to the dipole moment from their inter-layer component.  

In this Letter, we study the nature of the many-body states of {\it polaritons} formed by combining these strongly interacting hybrid excitons with a microcavity. We show that the very strong optical non-linearities arising from the interactions between the hybrid excitons can lead to features with no  parallel in conventional two-dimensional polariton gases based on semiconductor quantum wells. When the cavity is driven coherently, the steady-state exhibits a bi-stability which, in contrast to two-dimensional polariton gases~\cite{Carusotto2013}, is modulated by a discretised pattern, reminiscent of the equilibrium Mott insulating phase. When the excitons are pumped incoherently, we demonstrate that the strong correlations can cause the system to act as a  {\it multi-photon} laser, akin to that  of Rydberg and multi-level atoms~\cite{Brune1987,Davidovich1987,Lewenstein1990,Orszag1993}. 
 \begin{figure}
\begin{center}
\includegraphics[width=1\columnwidth]{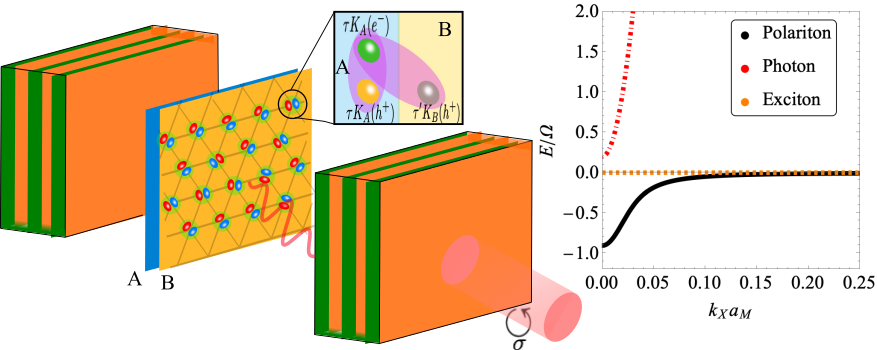}
\end{center}
\caption{(a) Schematic representation of spatially localised hybrid excitons embedded in a microcavity. (b) Zoom of typical polariton dispersions, for reference we show the photon and exciton dispersions, here we take $a_M=5\text{nm}$, $\Omega=15\text{meV}$ and $m_c=5\times 10^{-5}m_e$, with $m_e$ the bare electron mass. }
\label{Fig1}
\end{figure}  
%
In conventional semiconductor microcavities, most features -- including the experimental realisation of weakly interacting Bose-Einstein condensates~\cite{Kasprzak2006,Deng2010,Carusotto2013}, superfluid phases~\cite{Amo2009}, and quantum vortices~\cite{Dominici2018} of polaritons -- can be understood in terms of weakly interacting polaritons described by the Gross-Pitaevskii equation, or other non-linear classical description. Strong polariton-polariton interactions have been recently studied mainly in the context of Feshbach polaritonic resonances~\cite{Takemura2014,Sidler2017}, trion-mediated~\cite{Emmanuele2020}, and medium-induced optical non-linearities~\cite{Tan2020,Camacho2021}. 
However, to account for the physics of strongly interacting hybrid excitons in TMD bilayers~\cite{Zhang2021}, we must go beyond these works to consider the steady state properties of
a Hubbard model of excitons coupled to a high-finesse cavity. 

{\it Model.-} Consider a van der Waals heterostructure formed by two  semiconductor monolayers, labelled $A$ and $B$,   coupled to a microcavity as shown in Fig.~\ref{Fig1}(a). For TMD materials, the band extrema form two valleys, at wavevectors which we denote $\pm K_{A,B}$ for the two layers. For small twist angles, $K_{A} \simeq K_{B}$, the intra-layer excitons in which both electron and hole have valley index  $\tau  K_{A}$ $(\tau=\pm 1)$  hybridize preferentially with inter-layer excitons formed by $\tau K_{A}$ electrons and $\tau ' K_{B}$ holes with $\tau=\tau'$~\cite{Ruiz2019}. We consider the properties of the lowest energy band of the resulting hybrid excitons, and will focus attention on a single component $\tau=+1$, assuming that the light has circular polarization so drives only one valley.  The properties of this band of excitons can be well described by a two-dimensional Bose-Hubbard Hamiltonian
\begin{gather}
\hat H_X=\sum_{i}\omega_X\hat x^\dagger_i\hat x_i+\sum_i\frac{U_X}{2}\hat x^\dagger_i\hat x^\dagger_i\hat x_i\hat x_i-\sum_{\langle i,j\rangle} t_{ij}\hat x^\dagger_i\hat x_j,
\label{HX}
\end{gather}
where $\hat x^\dagger_i$ creates a $\tau = +1$  hybrid exciton in the Wannier orbital of the lowest band site $i$, which for now on we simply refer to as an exciton.  The on-site energy is given by $\omega_X,$ and the sites are arranged in a triangular lattice, connected by the tunnelling coefficients $t_{ij}$. The resulting band dispersion will not play a significant role in our results, since the local and repulsive interaction $U_X$ will be assumed to be much larger than $t_{ij}$, as is typical for experimental systems~\cite{Alexeev2019}. The small effects of exciton tunneling are discussed in detail in the Supplemental Material~\cite{SM}. Here, for clarity we set $t_{ij}=0.$


The light-matter coupling is described by
$\sum_{\mathbf k}\Omega\left(\hat a_{\mathbf k}^\dagger \hat x_{\mathbf k}+\hat x^\dagger_{\mathbf k}\hat a_{\mathbf k}\right)
$, 
where $\hat a^\dagger_{\mathbf k}$ ($\hat{x}^\dagger_{\mathbf k}$) creates a cavity photon (exciton) with in-plane momentum $\mathbf k$. (The photon is assumed to be circularly  polarised, to couple to the $\tau=+1$ valley exciton.) Here $\Omega$ is the Rabi coupling. 
The dispersion of free photons is $\hat H_l=\sum_{\mathbf k}\omega_c(\mathbf k)\hat a^\dagger_{\mathbf k}\hat a_{\mathbf k}$, with $\omega_c(\mathbf k)=\omega_c+|\mathbf k|^2/(2m_c),$ where $m_c$ is the cavity photon mass. Due to the steep dispersion of the light, for $\omega_X = \omega_c$, photons are decoupled from the excitons for momentum $\Delta k\approx\sqrt{2m_c\Omega}.$ For typical lattice constants $a_M\approx 1-10\text{nm},$ and Rabi coupling $\Omega \approx 10\text{meV}$, one finds $\Delta ka_M\approx 0.02-0.1\ll 1,$ so photons only couple to excitons within a very narrow region in the reduced Brillouin zone. As shown in Fig.~\ref{Fig1}(b), this introduces a length scale $\lambda$ for which excitons distanced by $\lambda/a_M=2\pi/(\Delta ka_M)\approx 60-300$  sites couple to the same $\mathbf k=0$ mode.  We therefore make the approximation that coupling is to a single mode, which we hereafter describe by $\hat{a}^{(\dag)}$ without momentum subscript, and take the cavity energy and light-matter interaction to be:
\begin{gather}
\hat H_{l} + \hat H_{l-m} = \omega_c \hat{a}^\dag \hat{a} + \frac{1}{\sqrt{N}}\sum_{ i}\Omega\left(\hat a^\dagger \hat x_{i}+\hat a \hat x_{i}^\dagger \right)\,.
\end{gather}

 Finally, in contrast to mobile two-dimensional excitons, saturation effects can become relevant even at the level of a few excitons per site, underlining the quasi-zero dimensional nature of the local exciton states. We account for this effect by adding an anharmonic light-matter coupling $\Omega(\hat n_i)=\Omega-\Omega_{\text{sat}}\hat n_i,$ which prevents an arbitrarily large excitation number $\hat{n}_i = \hat{x}^\dag_i\hat{x}_i$. Saturation effects are reminder of the strictly non-bosonic nature of the excitons, relevant when the inter-particle distance between the excitons is comparable to the exciton Bohr radius.

 We account for  intrinsic dissipative effects by studying the density matrix of the system,  $\hat\rho$,
 whose time evolution is governed by the Lindblad
 master equation~\cite{Carusotto2013}
\begin{gather}
\frac{d\hat\rho}{dt}=-i[\hat H,\hat\rho]+\mathcal D[\hat\rho]=\mathcal L[\hat\rho],
\label{ME}
\end{gather}
where $\hat H=\hat H_x+\hat H_l+\hat H_{l-m}$.
 The dissipative nature of the cavity photons and the excitons are represented via the Lindblad operator 
  \begin{gather}
\mathcal D[\hat\rho]=\frac{\gamma_c}{2}\left[2\hat a\hat\rho\hat a^\dagger-\{\hat a^\dagger\hat a,\hat\rho\}\right]+\sum_{i}\frac{\gamma_x}{2}\left[2\hat x_i\hat\rho\hat x^\dagger_i-\{\hat x^\dagger_i\hat x_i,\hat\rho\}\right],
\label{eq:excitonandcavitydecay}
\end{gather}
where $\gamma_c$ and $\gamma_x$ are the damping rates of the photons and excitons respectively. (We will later also introduce an incoherent pump of the exciton states.) 

In the field of cold atomic gases, Bose-Hubbard models coupled to cavity fields have been  studied in the context of optical lattices ~\cite{Brennecke2007,Ritsch2013,Mekhov2012}, which has led to experimental breakthroughs including the realisation of the Dicke phase transition~\cite{Baumann2010} and supersolidity~\cite{Leonard2017}. Additionally, Bose condensation of polaritons has been considered for spatially localised two-level systems coupled to a microcavity within the Dicke model~\cite{Eastham2001,Szymanska2003,Keeling2004}. The model we study, Eq.~\ref{HX}, goes beyond this by allowing multiple excitons per site, and the Hubbard interaction $U_X$ will play a key role.  Despite their intriguing potential to realise complex quantum phases of light and matter, the study of many-body phases of spatially localised  excitons in van der Waals heterostructures coupled to microcavities remains largely unexplored. Here, we exploit the unique features of the excitons to generate one- and multi- photon resonances by either injection of coherent photons and incoherent driving of excitons. 


{\it Coherent photon driving and bistability -} We start by considering a coherent drive of the cavity, as described by the term $(F\hat a^\dagger e^{-i\omega_p t}+F^*\hat a e^{i\omega_p t})$ with an amplitude $F$ and frequency  $\omega_p$. To determine the steady-state of the system we employ on-site exact-diagonalisation for the excitons combined with a self-consistent mean-field approach for the cavity photons. Since the optical non-linearities arise indirectly from the exciton-exciton interactions, we treat the excitons at the full quantum level and obtain recursively the steady-state properties of the cavity field at the semi-classical level. This goes beyond from an extended Gross-Pitaevskii for the compound object of exciton-photon in terms of polaritons~\cite{Carusotto2013}. The cavity field is given then by $\langle \hat a\rangle=\sqrt{N}\alpha,$ the amplitude $\alpha$ follows for the steady-state $\alpha=\frac{1}{\Delta_c}(f+\Omega\langle x\rangle),$ where $f=F/\sqrt{N}$ and we have taken $\sum_{i}\langle \hat x_i\rangle/N=\langle x\rangle$. In addition, we introduce $\Delta_c=\Delta\omega_c+i\gamma_c/2$ with $\Delta\omega_c=\omega_p-\omega_c$, the photon detuning. This defines an effective Hamiltonian for the excitons where the photon field is replaced by the amplitude $\sqrt{N}\alpha$.
The steady-state of the system is obtained by  exact diagonalisation of the super-operator $\mathcal L[\hat\rho]=-i[\hat H_{\text{local}},\hat\rho]+\mathcal D[\hat\rho]$ where $\hat H_{\text{local}}$ is the effective exciton Hamiltonian with the cavity field $\alpha$ a parameter obtained self-consistently, as explained in SM~\cite{SM}.

For clarity, consider  first  the regime of negligible saturation effects, $\Omega_{\text{sat}}/\Omega=0$.  In this case, the exciton Hamiltonian takes the form 
\begin{gather}
\hat H_{\text{local}}=\hat H_{X}+\sum_i(f_X\hat x_i^\dagger+f_X^*\hat x_i),
\end{gather} which introduces an effective driving
$f_X=\frac{\Omega\,f}{\Delta_c}+\left(\frac{\Omega^2}{\Delta_c}\right)\langle \hat x\rangle
$.
Within this regime ($\Omega_{\text{sat}}=0)$ the density matrix can also be obtained by a self-consistent approach based on an analytical expression for the density matrix, see~\cite{SM,Drummond1980}.

For exciton detuning $\Delta \omega_X = \omega_p - \omega_X> 0$, the onsite repulsion $U_X$ can bring the transition into resonance as the density of excitons increases, leading to a sudden increase in exciton density. This effect gives rise to a bistability as a function of $f/U_X$. In Fig.~\ref{Fig2} (b) we show the mean exciton density for increasing $f/U_X,$ that is when $f/U_X$ is varied from below. On the other hand,  Fig.~\ref{Fig2} (c) corresponds to the hysteresis branch, that is, a decreasing $f/U_X$ from an initial large positive $f/U_X$, here the steady-state remains in a high-density phase for smaller values of $f/U_X$ compared to Fig.~\ref{Fig2} (b) indicating the hysteresis. The mismatch between these plots constructs the phase-diagram in Fig.~\ref{Fig2}(a). The phase-diagram consists of two monostable phases separated by a bi-stability. 
\begin{figure}[h!]
\begin{center}
\includegraphics[width=1\columnwidth]{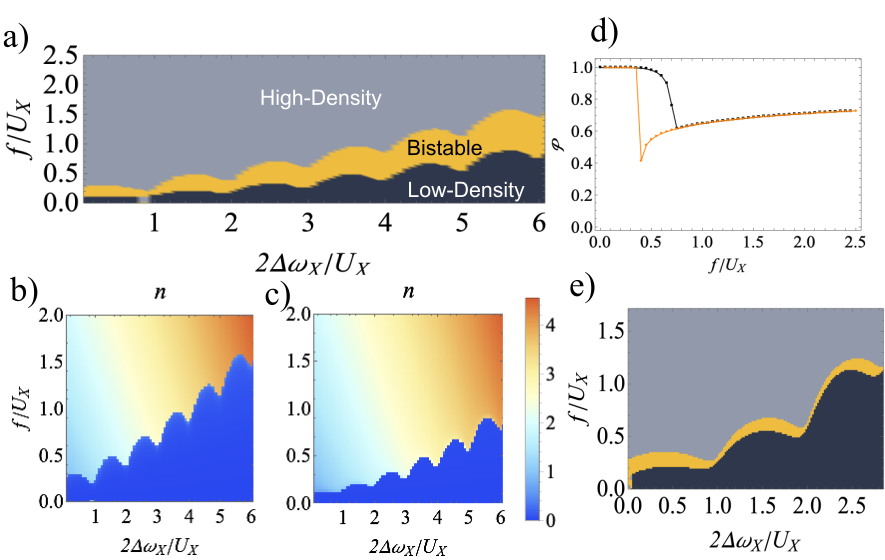}
\end{center}
\caption{(a) Stability phase diagram: black/grey regions correspond to low/high-density phases, separated by a  bi-stability region (orange). Exciton number as a function of  $f/U_X$ and $\Delta\omega_X/U_X$, at $\Delta\omega_c/U_X=-1$, showing hysteresis for: (b) increasing $f/U_X$; and (c) decreasing $f/U_X$.  (d) Purity of the steady state at   $\Delta\omega_X/U_X=1.2$, for the two hysteretic branches, of increasing/decreasing $f/U_X$ (black/orange). (e) Stability phase diagram showing the effects of saturation, with $\Omega_{\text{sat}}=\Omega/4$. }
\label{Fig2}
\end{figure}  
The distinctive lobular pattern in the phase diagram is determined by the $n$-exciton resonance,
\begin{gather}
(n-1)=\frac{2\Delta\omega_X}{U_X}.
\label{eR}
\end{gather}
Physically, the transition to the high-density phase with exciton number $n$ can be understood as an energetic condition. When the energy of $n$-incoming photons equals the energy of $n$-interacting excitons $n\omega_p=n\omega_X+n(n-1)U_X/2$ this transition is enhanced.
We illustrate our results for physically motivated parameters, we fix $\gamma_x/U_X=0.2$, $\gamma_c/U_X=0.2,$ and $\Delta\omega_c/U_X=-1$. 

The bi-stable phase is related to a tunnelling-mediated bi-stability, where the light-matter coupling induces effective exciton hopping through the lattice with amplitude $J_{\text{med}}\propto -\Omega^2f/\Delta_c.$ This is a second-order process, where an exciton emits a photon which decays creating an exciton in a different site. 
A similar bistability arises within a mean-field treatment of photonic lattices, as a consequence of photon hopping between the cavities~\cite{Drummond1980,Boite2013,Boite2014,Biondi2017}. We emphasize that the Bose-Hubbard-cavity model we study differs from these models of photonic cavities through the existence of both excitonic and photonic degrees of freedom. Furthermore,   the parameter $\Delta k a_M \ll 1$ motivates the mean-field character of the collective coupling among multiple excitonic sites.

As highlighted above, for the electronic system one should also allow for saturation effects in Rabi coupling, which 
arise when the extension of the spatial localisation of the excitons is of the order of the exciton Bohr radius. For state-of-the-art experiments with moir\'e excitons  the extent of the exciton wave-function is of the order $r_s\approx 1-5\text{nm}$ with typical Bohr radius of $a_B\approx 1\text{nm}$~\cite{Alexeev2019,Zhang2021}. Roughly, one can estimate $\Omega_{\text{sat}}/\Omega\propto (a_B/r_s)^2\in \{0.1,1\},$ see details in~\cite{SM}. In Fig.~\ref{Fig2}(e) we illustrate for $\Omega_{\text{sat}}=\Omega/4$ the imprints of the anharmonic light-matter coupling on the phase diagram, with all other parameters as in Fig.~\ref{Fig2}(a).   For small exciton detuning, the lobular pattern respects the $n$- exciton resonances. Moreover, a bi-stable phase remainder of the phase-diagram in Fig.~\ref{Fig2}(a) persists within the lobes in Fig.~\ref{Fig2}(e). The region of bi-stability shrinks somewhat as saturation effects increase, but the cusps remain very prominent. 
%
Both the bi-stability and the cusps associated with the $n$-exciton resonances remain for saturation effects of the size expected.

%

The results that we find arise from quantum correlations that go beyond what can be obtained in a Gross-Pitaevskii (GP) mean field theory, which is appropriate for weakly interacting excitons. In the GP method, the excitons are in a pure quantum state -- albeit a coherent state with non-zero $\langle \hat{x}_i\rangle$. The states arising here are, in general, not pure. 
For example, the low- to high- density transition is accompanied by a change in the statistical nature of the density matrix, which evolves from a pure quantum state into a statistical mixture.  Fig.~\ref{Fig2} (d) shows the drop of the purity  $
\mathcal P=\text{Tr}(\hat\rho^2),$ for  $\Delta\omega_X/U_X=1.2$ along the transition.
Furthermore, for positive cavity detuning, the bi-stability is inhibited, and instead the transition from the low- to high-density regime is a smooth function of $f/U_X$. In Fig.~\ref{Fig4} we show  $|\langle \hat x\rangle|$ as a function of $f/U_X$ for $\Delta\omega_X/U_X=1.2,$ and $\Delta\omega_c/U_X=1$. To illustrate the distinct character of the localised excitons compared to their mobile two-dimensional counterpart, we compare the solutions obtained from the master equation, the extended Gross-Pitaevskii equation(eGP)~\cite{Carusotto2013}, and  non-interacting excitons, which for weak driving intensities $f/U_X,$ agree  well. However, when the exciton density is no longer small,  the failure of the eGP to capture the underlying excitation spectrum of Eq.~\ref{HX} becomes evident. On the one hand, the eGP predicts always an S-shaped bi-stability for $\Delta\omega_X>0$, a characteristic signature of weakly interacting polariton gases~\cite{Ciuti2005,Carusotto2013}. However, the steady-state finds no bi-stable phase and is characterised instead by an abrupt but continuous drop of the coherence at $f/U_X\approx 0.25$. In this regime, the results of the eGP  disagrees with those of the master equation approach both quantitatively and qualitatively. Figure~\ref{Fig4} highlights that the eGP sufficient for weakly interacting two-dimensional polariton gases, is no longer adequate to describe the coupling of localised excitons to light.
\begin{figure}[t]
\begin{center}
\includegraphics[width=\columnwidth]{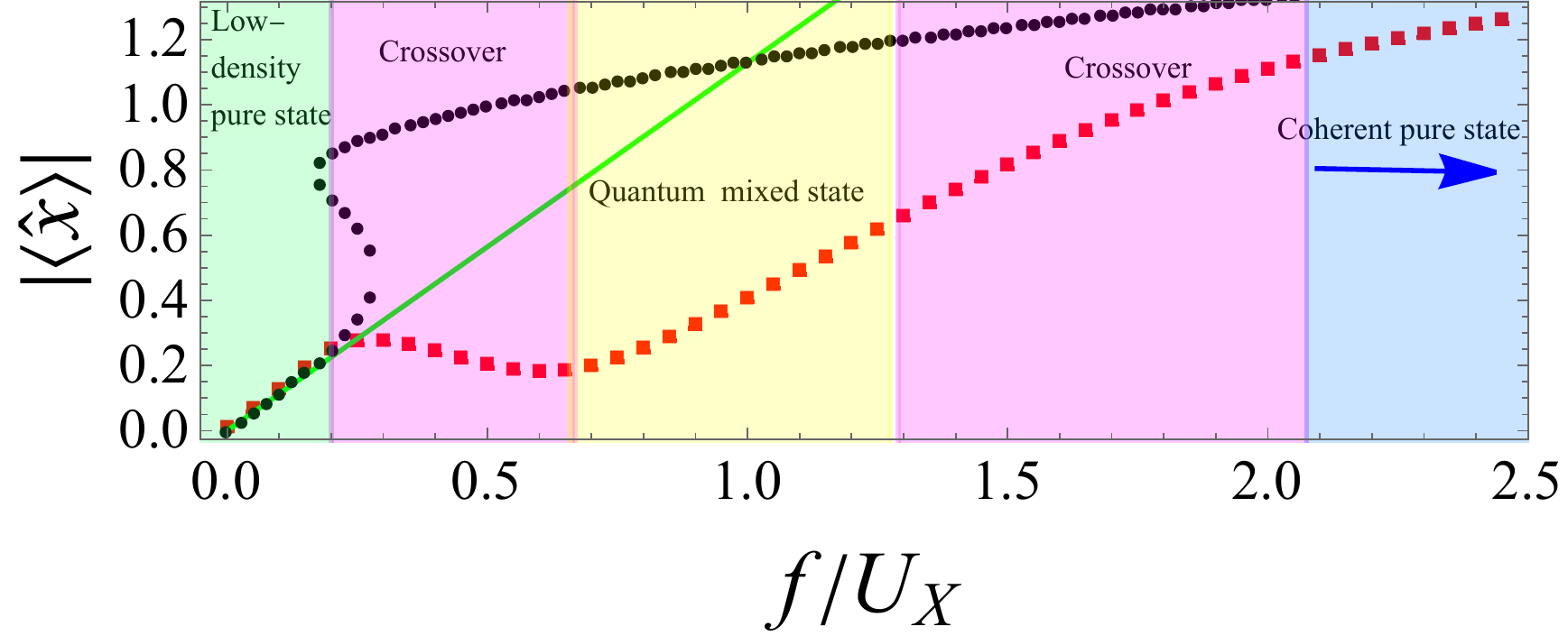}
\end{center}
\caption{Exciton amplitude $|\langle\hat x\rangle|$ for $\Delta\omega_X/U_X=1.2,$ and  $\Delta\omega_c/U_X=1.$ We show the eGP (black), self-consistent on-site diagonalization (red), and non-interacting excitons (green).} 
\label{Fig4}
\end{figure}

{\it Incoherent pumping and lasing.-} 
Lasers based on two-dimensional TMD's materials have attracted much the attention~\cite{Wen2020}, mostly in the context of photonic crystals~\cite{Wu2015}, whispering gallery microcavities~\cite{Ye2015,Salehzadeh2015}, polariton lasing~\cite{Shan2021,Carusotto2013}, and trion- induced optical gain~\cite{Tan2020,Wasak2021}. Interlayer excitons periodically trapped, however, unfold a promising new avenue, in which the inherited valley physics, strong interactions, and intra-inter exciton mixing make these systems ideal platforms to realise one- and multi-photon lasers. By optically driving higher excitonic bands which quickly relax to the hybrid exciton state, it is possible to engineer driving schemes with incoherent exciton gain~\cite{Paik2019}. In this case, we add to the exciton and cavity decay (\ref{eq:excitonandcavitydecay}) the term
\begin{gather}
\mathcal D_{\rm gain}[\hat\rho]=\frac{\Gamma_x}{2}\sum_{i}\left[2\hat y_i^\dagger\hat\rho\hat y_i-\{\hat y_i\hat y^\dagger_i,\hat\rho\}\right],
\end{gather}
which describes incoherent exciton gain at a rate $\Gamma_x$. We assume that the gain mechanism pumps the exciton number on each site up to an occupancy of at most $n_{\rm max}$. The jump operator is then $\hat y_i^{\dagger} \equiv \hat{x}_i \hat{P}_i,$  where $\hat{P}_i$ is the projection operator on site $i$ onto the exciton number states in the range $n_i=0,1\ldots n_{\rm max}-1$. 
For clarity, here we take $n_{\rm max}=2$.  The photon energy $\omega$ is calculated self-consistently to determine the steady-state, and it does not necessarily match the cavity $\omega_c$ nor the exciton $\omega_X$ energies~\cite{SM}. This energy is obtained by maintaining the semi-classical treatment for the photon field. Again, we assume that only one valley exciton is driven, for example by pumping with circular polarized light.

We find that a non-trivial state with $\alpha\neq 0$  emerges consequence of two processes, see Fig.~\ref{Fig6} (inset, right): (a) Single-photon resonances, when the energy of a single photon matches the energy of the transitions $|0\rangle\rightarrow|1\rangle$ and  $|1\rangle\rightarrow |2\rangle,$ with  $\omega\approx\omega_X$ and $\omega\approx\omega_X+U_X$ respectively, and (b) two-photon resonances corresponding to the emission of two-photons with $2\omega\approx2\omega_X+U_X.$

\begin{figure}[h!]
\begin{center}
\includegraphics[width=\columnwidth]{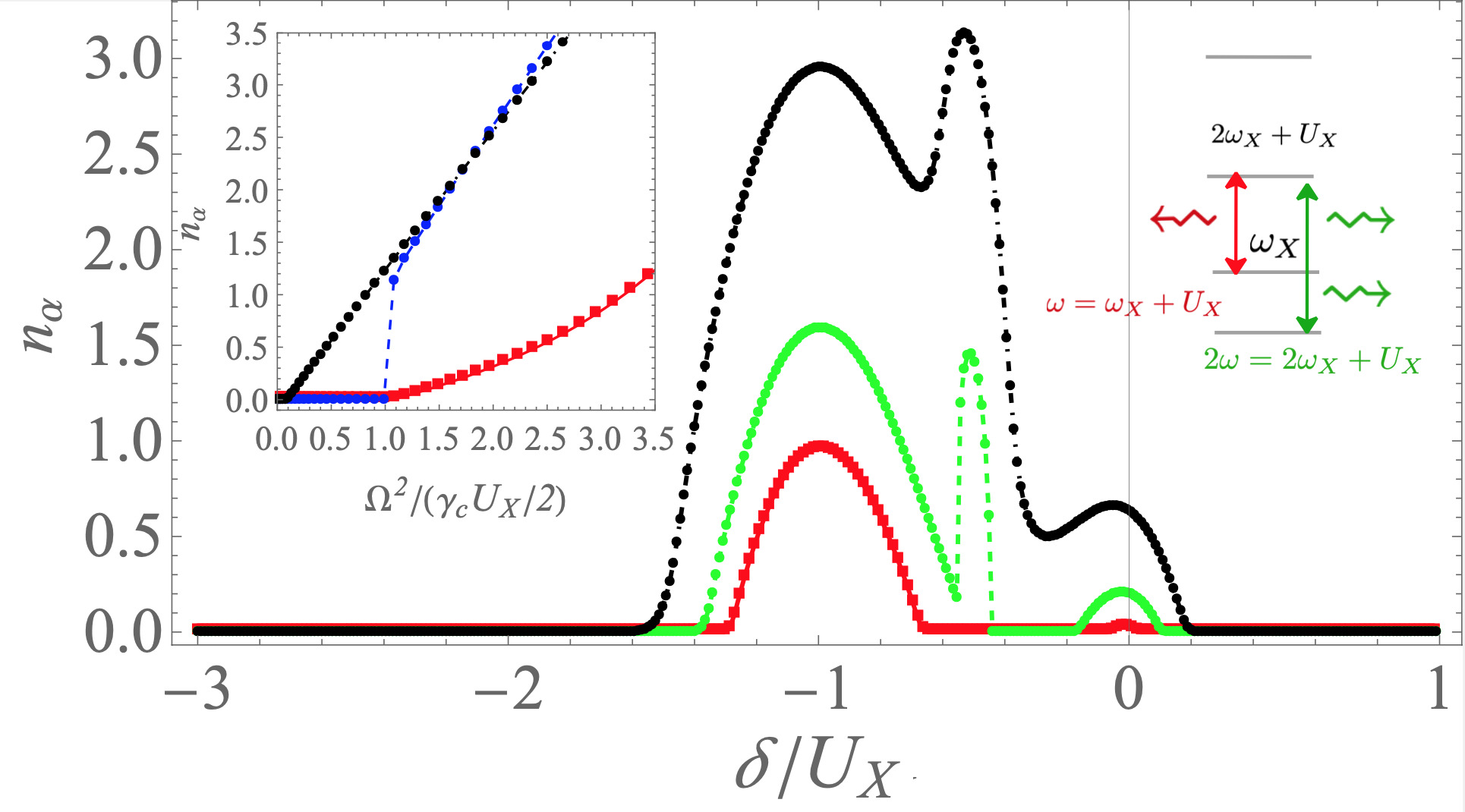}
\end{center}
\caption{ Photon amplitude $|\alpha|^2$ as a function of the cavity detuning for $\Omega/U_X=1/12$ (red), $\Omega/U_X=1/8$ (green), $\Omega/U_X=1/6$ (black), we fix $\Gamma_x/U=1/10$  $\Gamma_x/\gamma_X=6$, $\Gamma_x/\gamma_c=3$.  (Inset, left) Photon amplitude as a function of $2\Omega^2/(U_X\gamma_c)$ for varying $\Omega.$ Single-particle transitions $\delta/U_X=0$ (red) and $\delta/U_X=-1$ (black)   and $\delta/U_X=-1/2$ (blue). (Inset, right) Sketch of the exciton level scheme and the photon transitions, the wavy red arrow illustrates the one-photon resonance $|2\rangle\rightarrow|1\rangle$, while the green wavy arrows depict a two-photon process.  }
\label{Fig6}
\end{figure} 

{Figure~\ref{Fig6} shows the photon amplitude $|\alpha|^2$ for several values of the light-matter coupling $\Omega/U_X$ as a function of the cavity detuning $\delta=\omega_X-\omega_c$. The single-photon resonances can be understood in terms of effective two-level systems, where the photon energy $\omega\approx\omega_X$ or $\omega\approx\omega_X+U_X$ is resonant with the one-particle transitions $|0\rangle\rightarrow|1\rangle$  and  $|1\rangle\rightarrow |2\rangle,$ respectively. The injection of excitons induces population inversion, that is $\rho_{22}>\rho_{11}>\rho_{00},$ leading to an enhanced photon amplitude for the  $|1\rangle\rightarrow |2\rangle$ transition, compared to the one-photon transition with energy $\omega_c=\omega_X$. Interpreting the single-photon transitions in terms of effective two-level systems, the non-trivial solutions arise beyond a critical light-matter coupling  $\Omega_c=(\gamma_x+\Gamma_x)\sqrt{\frac{\gamma_c}{4(\Gamma_x-\gamma_x)}},$ see SM~\cite{SM}.

Emission of photons pairs is promoted at the two-photon resonance, that is, $\delta/U_X=-1/2,$ due to the strong exciton interactions: the single- and multi-photon resonances are well-separated in energy, and can be clearly distinguished in Fig.~\ref{Fig6}. Inspired by the studies on lasers based on two-photon gain in three-level atoms~\cite{Brune1987,Davidovich1987,Lewenstein1990,Orszag1993}, we introduce an effective two-photon coupling $g_2=\Omega^2/\Delta,$ where $\delta=U_X/2$ is fixed by the discrete levels of the excitonic spectrum. In Fig.~\ref{Fig6} (inset, left) we plot the amplitude of the cavity field as a function of $g_2/\gamma_c$ at the single-particle transitions $\delta/U_X=0$ (red) and $\delta/U_X=-1$ (black)  and at the two-photon resonance (blue). 
The sudden entrance of two-photon lasing can be also understood in terms of effective laser rate equations, which predict a photon number that discontinuously jumps to $n_\alpha\approx 2\Gamma_x/(5\gamma_c),$ with a two-photon stimulated rate of $A_2\propto g_2^2/\Gamma_x,$ see details in SM~\cite{SM}. Here, saturation effects for $\Omega_s=\Omega/4$ lead only to quantitative corrections, leaving our main conclusions valid.


{\it Conclusions.-} The advent of the moir\'e polaritons~\cite{Zhang2021} opens up the door to the study of strongly correlated phases of matter coupled to cavity photons, promising a new generation of hybrid states of light and matter. 
We have demonstrated that as a consequence of the competition between exciton interaction, saturation effects, non-equilibrium features, and strong light-matter interactions, a rich phase diagram arises with no counterpart in conventional two-dimensional polariton gases. In state-of-the-art systems, the expected optical bi-stabilities could be detected by measuring the transmitted spectrum in a similar fashion to polariton gases~\cite{Baas2004,Paraiso2010,Ouellet2017}. Detection of the cusps in the bistability would allow a direct measurement of local Hubbard interaction of excitons. 
These materials also permit the incoherent drive of excitons, for example by optically driving higher excitonic bands which quickly relax to the inter-layer exciton state~\cite{Paik2019}. We have shown that the strong interactions allow these systems to realize multi-photon lasers.  


{\it Acknowledgments.-} We are grateful to Mete Atat{\" u}re, Jeremy Baumberg and Benjamin Remez for helpful discussions. This work was supported by EPSRC Grant Nos. EP/P009565/1, EP/P034616/1 and by a Simons Investigator Award.

  \bibliography{moire}

\begin{thebibliography}{60}%
\makeatletter
\providecommand \@ifxundefined [1]{%
 \@ifx{#1\undefined}
}%
\providecommand \@ifnum [1]{%
 \ifnum #1\expandafter \@firstoftwo
 \else \expandafter \@secondoftwo
 \fi
}%
\providecommand \@ifx [1]{%
 \ifx #1\expandafter \@firstoftwo
 \else \expandafter \@secondoftwo
 \fi
}%
\providecommand \natexlab [1]{#1}%
\providecommand \enquote  [1]{``#1''}%
\providecommand \bibnamefont  [1]{#1}%
\providecommand \bibfnamefont [1]{#1}%
\providecommand \citenamefont [1]{#1}%
\providecommand \href@noop [0]{\@secondoftwo}%
\providecommand \href [0]{\begingroup \@sanitize@url \@href}%
\providecommand \@href[1]{\@@startlink{#1}\@@href}%
\providecommand \@@href[1]{\endgroup#1\@@endlink}%
\providecommand \@sanitize@url [0]{\catcode `\\12\catcode `\$12\catcode
  `\&12\catcode `\#12\catcode `\^12\catcode `\_12\catcode `\%12\relax}%
\providecommand \@@startlink[1]{}%
\providecommand \@@endlink[0]{}%
\providecommand \url  [0]{\begingroup\@sanitize@url \@url }%
\providecommand \@url [1]{\endgroup\@href {#1}{\urlprefix }}%
\providecommand \urlprefix  [0]{URL }%
\providecommand \Eprint [0]{\href }%
\providecommand \doibase [0]{http://dx.doi.org/}%
\providecommand \selectlanguage [0]{\@gobble}%
\providecommand \bibinfo  [0]{\@secondoftwo}%
\providecommand \bibfield  [0]{\@secondoftwo}%
\providecommand \translation [1]{[#1]}%
\providecommand \BibitemOpen [0]{}%
\providecommand \bibitemStop [0]{}%
\providecommand \bibitemNoStop [0]{.\EOS\space}%
\providecommand \EOS [0]{\spacefactor3000\relax}%
\providecommand \BibitemShut  [1]{\csname bibitem#1\endcsname}%
\let\auto@bib@innerbib\@empty
\bibitem [{\citenamefont {Manzeli}\ \emph {et~al.}(2017)\citenamefont
  {Manzeli}, \citenamefont {Ovchinnikov}, \citenamefont {Pasquier},
  \citenamefont {Yazyev},\ and\ \citenamefont {Kis}}]{Manzeli2017}%
  \BibitemOpen
  \bibfield  {author} {\bibinfo {author} {\bibfnamefont {S.}~\bibnamefont
  {Manzeli}}, \bibinfo {author} {\bibfnamefont {D.}~\bibnamefont
  {Ovchinnikov}}, \bibinfo {author} {\bibfnamefont {D.}~\bibnamefont
  {Pasquier}}, \bibinfo {author} {\bibfnamefont {O.~V.}\ \bibnamefont
  {Yazyev}}, \ and\ \bibinfo {author} {\bibfnamefont {A.}~\bibnamefont {Kis}},\
  }\href {\doibase 10.1038/natrevmats.2017.33} {\bibfield  {journal} {\bibinfo
  {journal} {Nature Reviews Materials}\ }\textbf {\bibinfo {volume} {2}},\
  \bibinfo {pages} {17033} (\bibinfo {year} {2017})}\BibitemShut {NoStop}%
\bibitem [{\citenamefont {Tang}\ \emph {et~al.}(2020)\citenamefont {Tang},
  \citenamefont {Li}, \citenamefont {Li}, \citenamefont {Xu}, \citenamefont
  {Liu}, \citenamefont {Barmak}, \citenamefont {Watanabe}, \citenamefont
  {Taniguchi}, \citenamefont {MacDonald}, \citenamefont {Shan},\ and\
  \citenamefont {Mak}}]{Tang2020}%
  \BibitemOpen
  \bibfield  {author} {\bibinfo {author} {\bibfnamefont {Y.}~\bibnamefont
  {Tang}}, \bibinfo {author} {\bibfnamefont {L.}~\bibnamefont {Li}}, \bibinfo
  {author} {\bibfnamefont {T.}~\bibnamefont {Li}}, \bibinfo {author}
  {\bibfnamefont {Y.}~\bibnamefont {Xu}}, \bibinfo {author} {\bibfnamefont
  {S.}~\bibnamefont {Liu}}, \bibinfo {author} {\bibfnamefont {K.}~\bibnamefont
  {Barmak}}, \bibinfo {author} {\bibfnamefont {K.}~\bibnamefont {Watanabe}},
  \bibinfo {author} {\bibfnamefont {T.}~\bibnamefont {Taniguchi}}, \bibinfo
  {author} {\bibfnamefont {A.~H.}\ \bibnamefont {MacDonald}}, \bibinfo {author}
  {\bibfnamefont {J.}~\bibnamefont {Shan}}, \ and\ \bibinfo {author}
  {\bibfnamefont {K.~F.}\ \bibnamefont {Mak}},\ }\href {\doibase
  10.1038/s41586-020-2085-3} {\bibfield  {journal} {\bibinfo  {journal}
  {Nature}\ }\textbf {\bibinfo {volume} {579}},\ \bibinfo {pages} {353}
  (\bibinfo {year} {2020})}\BibitemShut {NoStop}%
\bibitem [{\citenamefont {Kennes}\ \emph {et~al.}(2021)\citenamefont {Kennes},
  \citenamefont {Claassen}, \citenamefont {Xian}, \citenamefont {Georges},
  \citenamefont {Millis}, \citenamefont {Hone}, \citenamefont {Dean},
  \citenamefont {Basov}, \citenamefont {Pasupathy},\ and\ \citenamefont
  {Rubio}}]{Kennes2021}%
  \BibitemOpen
  \bibfield  {author} {\bibinfo {author} {\bibfnamefont {D.~M.}\ \bibnamefont
  {Kennes}}, \bibinfo {author} {\bibfnamefont {M.}~\bibnamefont {Claassen}},
  \bibinfo {author} {\bibfnamefont {L.}~\bibnamefont {Xian}}, \bibinfo {author}
  {\bibfnamefont {A.}~\bibnamefont {Georges}}, \bibinfo {author} {\bibfnamefont
  {A.~J.}\ \bibnamefont {Millis}}, \bibinfo {author} {\bibfnamefont
  {J.}~\bibnamefont {Hone}}, \bibinfo {author} {\bibfnamefont {C.~R.}\
  \bibnamefont {Dean}}, \bibinfo {author} {\bibfnamefont {D.~N.}\ \bibnamefont
  {Basov}}, \bibinfo {author} {\bibfnamefont {A.~N.}\ \bibnamefont
  {Pasupathy}}, \ and\ \bibinfo {author} {\bibfnamefont {A.}~\bibnamefont
  {Rubio}},\ }\href {\doibase 10.1038/s41567-020-01154-3} {\bibfield  {journal}
  {\bibinfo  {journal} {Nature Physics}\ }\textbf {\bibinfo {volume} {17}},\
  \bibinfo {pages} {155} (\bibinfo {year} {2021})}\BibitemShut {NoStop}%
\bibitem [{\citenamefont {Baek}\ \emph {et~al.}(2020)\citenamefont {Baek},
  \citenamefont {Brotons-Gisbert}, \citenamefont {Koong}, \citenamefont
  {Campbell}, \citenamefont {Rambach}, \citenamefont {Watanabe}, \citenamefont
  {Taniguchi},\ and\ \citenamefont {Gerardot}}]{Baek2020}%
  \BibitemOpen
  \bibfield  {author} {\bibinfo {author} {\bibfnamefont {H.}~\bibnamefont
  {Baek}}, \bibinfo {author} {\bibfnamefont {M.}~\bibnamefont
  {Brotons-Gisbert}}, \bibinfo {author} {\bibfnamefont {Z.~X.}\ \bibnamefont
  {Koong}}, \bibinfo {author} {\bibfnamefont {A.}~\bibnamefont {Campbell}},
  \bibinfo {author} {\bibfnamefont {M.}~\bibnamefont {Rambach}}, \bibinfo
  {author} {\bibfnamefont {K.}~\bibnamefont {Watanabe}}, \bibinfo {author}
  {\bibfnamefont {T.}~\bibnamefont {Taniguchi}}, \ and\ \bibinfo {author}
  {\bibfnamefont {B.~D.}\ \bibnamefont {Gerardot}},\ }\href {\doibase
  10.1126/sciadv.aba8526} {\bibfield  {journal} {\bibinfo  {journal} {Science
  Advances}\ }\textbf {\bibinfo {volume} {6}} (\bibinfo {year} {2020}),\
  10.1126/sciadv.aba8526}\BibitemShut {NoStop}%
\bibitem [{\citenamefont {Yu}\ and\ \citenamefont {Yao}(2020)}]{Yu2020}%
  \BibitemOpen
  \bibfield  {author} {\bibinfo {author} {\bibfnamefont {H.}~\bibnamefont
  {Yu}}\ and\ \bibinfo {author} {\bibfnamefont {W.}~\bibnamefont {Yao}},\
  }\href {\doibase https://doi.org/10.1016/j.scib.2020.05.030} {\bibfield
  {journal} {\bibinfo  {journal} {Science Bulletin}\ }\textbf {\bibinfo
  {volume} {65}},\ \bibinfo {pages} {1555} (\bibinfo {year}
  {2020})}\BibitemShut {NoStop}%
\bibitem [{\citenamefont {Liu}\ \emph {et~al.}(2016)\citenamefont {Liu},
  \citenamefont {Weiss}, \citenamefont {Duan}, \citenamefont {Cheng},
  \citenamefont {Huang},\ and\ \citenamefont {Duan}}]{Liu2016}%
  \BibitemOpen
  \bibfield  {author} {\bibinfo {author} {\bibfnamefont {Y.}~\bibnamefont
  {Liu}}, \bibinfo {author} {\bibfnamefont {N.~O.}\ \bibnamefont {Weiss}},
  \bibinfo {author} {\bibfnamefont {X.}~\bibnamefont {Duan}}, \bibinfo {author}
  {\bibfnamefont {H.-C.}\ \bibnamefont {Cheng}}, \bibinfo {author}
  {\bibfnamefont {Y.}~\bibnamefont {Huang}}, \ and\ \bibinfo {author}
  {\bibfnamefont {X.}~\bibnamefont {Duan}},\ }\href {\doibase
  10.1038/natrevmats.2016.42} {\bibfield  {journal} {\bibinfo  {journal}
  {Nature Reviews Materials}\ }\textbf {\bibinfo {volume} {1}},\ \bibinfo
  {pages} {16042} (\bibinfo {year} {2016})}\BibitemShut {NoStop}%
\bibitem [{\citenamefont {Schaibley}\ \emph {et~al.}(2016)\citenamefont
  {Schaibley}, \citenamefont {Yu}, \citenamefont {Clark}, \citenamefont
  {Rivera}, \citenamefont {Ross}, \citenamefont {Seyler}, \citenamefont {Yao},\
  and\ \citenamefont {Xu}}]{Schaibley2016}%
  \BibitemOpen
  \bibfield  {author} {\bibinfo {author} {\bibfnamefont {J.~R.}\ \bibnamefont
  {Schaibley}}, \bibinfo {author} {\bibfnamefont {H.}~\bibnamefont {Yu}},
  \bibinfo {author} {\bibfnamefont {G.}~\bibnamefont {Clark}}, \bibinfo
  {author} {\bibfnamefont {P.}~\bibnamefont {Rivera}}, \bibinfo {author}
  {\bibfnamefont {J.~S.}\ \bibnamefont {Ross}}, \bibinfo {author}
  {\bibfnamefont {K.~L.}\ \bibnamefont {Seyler}}, \bibinfo {author}
  {\bibfnamefont {W.}~\bibnamefont {Yao}}, \ and\ \bibinfo {author}
  {\bibfnamefont {X.}~\bibnamefont {Xu}},\ }\href {\doibase
  10.1038/natrevmats.2016.55} {\bibfield  {journal} {\bibinfo  {journal}
  {Nature Reviews Materials}\ }\textbf {\bibinfo {volume} {1}},\ \bibinfo
  {pages} {16055} (\bibinfo {year} {2016})}\BibitemShut {NoStop}%
\bibitem [{\citenamefont {Rasmita}\ and\ \citenamefont
  {Gao}(2020)}]{Rasmita2020}%
  \BibitemOpen
  \bibfield  {author} {\bibinfo {author} {\bibfnamefont {A.}~\bibnamefont
  {Rasmita}}\ and\ \bibinfo {author} {\bibfnamefont {W.-b.}\ \bibnamefont
  {Gao}},\ }\href {\doibase 10.1007/s12274-020-3036-x} {\bibfield  {journal}
  {\bibinfo  {journal} {Nano Research}\ } (\bibinfo {year} {2020}),\
  10.1007/s12274-020-3036-x}\BibitemShut {NoStop}%
\bibitem [{\citenamefont {Tang}\ \emph {et~al.}(2021)\citenamefont {Tang},
  \citenamefont {Gu}, \citenamefont {Liu}, \citenamefont {Watanabe},
  \citenamefont {Taniguchi}, \citenamefont {Hone}, \citenamefont {Mak},\ and\
  \citenamefont {Shan}}]{Tang2020a}%
  \BibitemOpen
  \bibfield  {author} {\bibinfo {author} {\bibfnamefont {Y.}~\bibnamefont
  {Tang}}, \bibinfo {author} {\bibfnamefont {J.}~\bibnamefont {Gu}}, \bibinfo
  {author} {\bibfnamefont {S.}~\bibnamefont {Liu}}, \bibinfo {author}
  {\bibfnamefont {K.}~\bibnamefont {Watanabe}}, \bibinfo {author}
  {\bibfnamefont {T.}~\bibnamefont {Taniguchi}}, \bibinfo {author}
  {\bibfnamefont {J.}~\bibnamefont {Hone}}, \bibinfo {author} {\bibfnamefont
  {K.~F.}\ \bibnamefont {Mak}}, \ and\ \bibinfo {author} {\bibfnamefont
  {J.}~\bibnamefont {Shan}},\ }\href {\doibase 10.1038/s41565-020-00783-2}
  {\bibfield  {journal} {\bibinfo  {journal} {Nature Nanotechnology}\ }\textbf
  {\bibinfo {volume} {16}},\ \bibinfo {pages} {52} (\bibinfo {year}
  {2021})}\BibitemShut {NoStop}%
\bibitem [{\citenamefont {F{\"o}rg}\ \emph {et~al.}(2021)\citenamefont
  {F{\"o}rg}, \citenamefont {Baimuratov}, \citenamefont {Kruchinin},
  \citenamefont {Vovk}, \citenamefont {Scherzer}, \citenamefont {F{\"o}rste},
  \citenamefont {Funk}, \citenamefont {Watanabe}, \citenamefont {Taniguchi},\
  and\ \citenamefont {H{\"o}gele}}]{Forg2021}%
  \BibitemOpen
  \bibfield  {author} {\bibinfo {author} {\bibfnamefont {M.}~\bibnamefont
  {F{\"o}rg}}, \bibinfo {author} {\bibfnamefont {A.~S.}\ \bibnamefont
  {Baimuratov}}, \bibinfo {author} {\bibfnamefont {S.~Y.}\ \bibnamefont
  {Kruchinin}}, \bibinfo {author} {\bibfnamefont {I.~A.}\ \bibnamefont {Vovk}},
  \bibinfo {author} {\bibfnamefont {J.}~\bibnamefont {Scherzer}}, \bibinfo
  {author} {\bibfnamefont {J.}~\bibnamefont {F{\"o}rste}}, \bibinfo {author}
  {\bibfnamefont {V.}~\bibnamefont {Funk}}, \bibinfo {author} {\bibfnamefont
  {K.}~\bibnamefont {Watanabe}}, \bibinfo {author} {\bibfnamefont
  {T.}~\bibnamefont {Taniguchi}}, \ and\ \bibinfo {author} {\bibfnamefont
  {A.}~\bibnamefont {H{\"o}gele}},\ }\href {\doibase
  10.1038/s41467-021-21822-z} {\bibfield  {journal} {\bibinfo  {journal}
  {Nature Communications}\ }\textbf {\bibinfo {volume} {12}},\ \bibinfo {pages}
  {1656} (\bibinfo {year} {2021})}\BibitemShut {NoStop}%
\bibitem [{\citenamefont {Hsu}\ \emph {et~al.}(2019)\citenamefont {Hsu},
  \citenamefont {Lin}, \citenamefont {Lu}, \citenamefont {Lee}, \citenamefont
  {Chu}, \citenamefont {Li}, \citenamefont {Yao}, \citenamefont {Chang},\ and\
  \citenamefont {Shih}}]{Hsu2019}%
  \BibitemOpen
  \bibfield  {author} {\bibinfo {author} {\bibfnamefont {W.-T.}\ \bibnamefont
  {Hsu}}, \bibinfo {author} {\bibfnamefont {B.-H.}\ \bibnamefont {Lin}},
  \bibinfo {author} {\bibfnamefont {L.-S.}\ \bibnamefont {Lu}}, \bibinfo
  {author} {\bibfnamefont {M.-H.}\ \bibnamefont {Lee}}, \bibinfo {author}
  {\bibfnamefont {M.-W.}\ \bibnamefont {Chu}}, \bibinfo {author} {\bibfnamefont
  {L.-J.}\ \bibnamefont {Li}}, \bibinfo {author} {\bibfnamefont
  {W.}~\bibnamefont {Yao}}, \bibinfo {author} {\bibfnamefont {W.-H.}\
  \bibnamefont {Chang}}, \ and\ \bibinfo {author} {\bibfnamefont {C.-K.}\
  \bibnamefont {Shih}},\ }\href {\doibase 10.1126/sciadv.aax7407} {\bibfield
  {journal} {\bibinfo  {journal} {Science Advances}\ }\textbf {\bibinfo
  {volume} {5}} (\bibinfo {year} {2019}),\ 10.1126/sciadv.aax7407}\BibitemShut
  {NoStop}%
\bibitem [{\citenamefont {Jung}\ \emph {et~al.}(2014)\citenamefont {Jung},
  \citenamefont {Raoux}, \citenamefont {Qiao},\ and\ \citenamefont
  {MacDonald}}]{Jung2014}%
  \BibitemOpen
  \bibfield  {author} {\bibinfo {author} {\bibfnamefont {J.}~\bibnamefont
  {Jung}}, \bibinfo {author} {\bibfnamefont {A.}~\bibnamefont {Raoux}},
  \bibinfo {author} {\bibfnamefont {Z.}~\bibnamefont {Qiao}}, \ and\ \bibinfo
  {author} {\bibfnamefont {A.~H.}\ \bibnamefont {MacDonald}},\ }\href {\doibase
  10.1103/PhysRevB.89.205414} {\bibfield  {journal} {\bibinfo  {journal} {Phys.
  Rev. B}\ }\textbf {\bibinfo {volume} {89}},\ \bibinfo {pages} {205414}
  (\bibinfo {year} {2014})}\BibitemShut {NoStop}%
\bibitem [{\citenamefont {Zhang}\ \emph {et~al.}(2017)\citenamefont {Zhang},
  \citenamefont {Chuu}, \citenamefont {Ren}, \citenamefont {Li}, \citenamefont
  {Li}, \citenamefont {Jin}, \citenamefont {Chou},\ and\ \citenamefont
  {Shih}}]{Zhang2017}%
  \BibitemOpen
  \bibfield  {author} {\bibinfo {author} {\bibfnamefont {C.}~\bibnamefont
  {Zhang}}, \bibinfo {author} {\bibfnamefont {C.-P.}\ \bibnamefont {Chuu}},
  \bibinfo {author} {\bibfnamefont {X.}~\bibnamefont {Ren}}, \bibinfo {author}
  {\bibfnamefont {M.-Y.}\ \bibnamefont {Li}}, \bibinfo {author} {\bibfnamefont
  {L.-J.}\ \bibnamefont {Li}}, \bibinfo {author} {\bibfnamefont
  {C.}~\bibnamefont {Jin}}, \bibinfo {author} {\bibfnamefont {M.-Y.}\
  \bibnamefont {Chou}}, \ and\ \bibinfo {author} {\bibfnamefont {C.-K.}\
  \bibnamefont {Shih}},\ }\href {\doibase 10.1126/sciadv.1601459} {\bibfield
  {journal} {\bibinfo  {journal} {Science Advances}\ }\textbf {\bibinfo
  {volume} {3}} (\bibinfo {year} {2017}),\ 10.1126/sciadv.1601459}\BibitemShut
  {NoStop}%
\bibitem [{\citenamefont {Ulstrup}\ \emph {et~al.}(2020)\citenamefont
  {Ulstrup}, \citenamefont {Koch}, \citenamefont {Singh}, \citenamefont
  {McCreary}, \citenamefont {Jonker}, \citenamefont {Robinson}, \citenamefont
  {Jozwiak}, \citenamefont {Rotenberg}, \citenamefont {Bostwick}, \citenamefont
  {Katoch},\ and\ \citenamefont {Miwa}}]{Ulstrup2020}%
  \BibitemOpen
  \bibfield  {author} {\bibinfo {author} {\bibfnamefont {S.}~\bibnamefont
  {Ulstrup}}, \bibinfo {author} {\bibfnamefont {R.~J.}\ \bibnamefont {Koch}},
  \bibinfo {author} {\bibfnamefont {S.}~\bibnamefont {Singh}}, \bibinfo
  {author} {\bibfnamefont {K.~M.}\ \bibnamefont {McCreary}}, \bibinfo {author}
  {\bibfnamefont {B.~T.}\ \bibnamefont {Jonker}}, \bibinfo {author}
  {\bibfnamefont {J.~T.}\ \bibnamefont {Robinson}}, \bibinfo {author}
  {\bibfnamefont {C.}~\bibnamefont {Jozwiak}}, \bibinfo {author} {\bibfnamefont
  {E.}~\bibnamefont {Rotenberg}}, \bibinfo {author} {\bibfnamefont
  {A.}~\bibnamefont {Bostwick}}, \bibinfo {author} {\bibfnamefont
  {J.}~\bibnamefont {Katoch}}, \ and\ \bibinfo {author} {\bibfnamefont {J.~A.}\
  \bibnamefont {Miwa}},\ }\href {\doibase 10.1126/sciadv.aay6104} {\bibfield
  {journal} {\bibinfo  {journal} {Science Advances}\ }\textbf {\bibinfo
  {volume} {6}} (\bibinfo {year} {2020}),\ 10.1126/sciadv.aay6104}\BibitemShut
  {NoStop}%
\bibitem [{\citenamefont {Tran}\ \emph {et~al.}(2020)\citenamefont {Tran},
  \citenamefont {Choi},\ and\ \citenamefont {Singh}}]{Tran2020}%
  \BibitemOpen
  \bibfield  {author} {\bibinfo {author} {\bibfnamefont {K.}~\bibnamefont
  {Tran}}, \bibinfo {author} {\bibfnamefont {J.}~\bibnamefont {Choi}}, \ and\
  \bibinfo {author} {\bibfnamefont {A.}~\bibnamefont {Singh}},\ }\href
  {\doibase 10.1088/2053-1583/abd3e7} {\bibfield  {journal} {\bibinfo
  {journal} {2D Materials}\ }\textbf {\bibinfo {volume} {8}},\ \bibinfo {pages}
  {022002} (\bibinfo {year} {2020})}\BibitemShut {NoStop}%
\bibitem [{\citenamefont {Jiang}\ \emph {et~al.}(2021)\citenamefont {Jiang},
  \citenamefont {Chen}, \citenamefont {Zheng}, \citenamefont {Zheng},\ and\
  \citenamefont {Pan}}]{Jiang2021}%
  \BibitemOpen
  \bibfield  {author} {\bibinfo {author} {\bibfnamefont {Y.}~\bibnamefont
  {Jiang}}, \bibinfo {author} {\bibfnamefont {S.}~\bibnamefont {Chen}},
  \bibinfo {author} {\bibfnamefont {W.}~\bibnamefont {Zheng}}, \bibinfo
  {author} {\bibfnamefont {B.}~\bibnamefont {Zheng}}, \ and\ \bibinfo {author}
  {\bibfnamefont {A.}~\bibnamefont {Pan}},\ }\href {\doibase
  10.1038/s41377-021-00500-1} {\bibfield  {journal} {\bibinfo  {journal}
  {Light: Science \& Applications}\ }\textbf {\bibinfo {volume} {10}},\
  \bibinfo {pages} {72} (\bibinfo {year} {2021})}\BibitemShut {NoStop}%
\bibitem [{\citenamefont {Tartakovskii}(2020)}]{Tartakovskii2020}%
  \BibitemOpen
  \bibfield  {author} {\bibinfo {author} {\bibfnamefont {A.}~\bibnamefont
  {Tartakovskii}},\ }\href {\doibase 10.1038/s42254-019-0136-1} {\bibfield
  {journal} {\bibinfo  {journal} {Nature Reviews Physics}\ }\textbf {\bibinfo
  {volume} {2}},\ \bibinfo {pages} {8} (\bibinfo {year} {2020})}\BibitemShut
  {NoStop}%
\bibitem [{\citenamefont {Alexeev}\ \emph {et~al.}(2019)\citenamefont
  {Alexeev}, \citenamefont {Ruiz-Tijerina}, \citenamefont {Danovich},
  \citenamefont {Hamer}, \citenamefont {Terry}, \citenamefont {Nayak},
  \citenamefont {Ahn}, \citenamefont {Pak}, \citenamefont {Lee}, \citenamefont
  {Sohn}, \citenamefont {Molas}, \citenamefont {Koperski}, \citenamefont
  {Watanabe}, \citenamefont {Taniguchi}, \citenamefont {Novoselov},
  \citenamefont {Gorbachev}, \citenamefont {Shin}, \citenamefont {Fal'ko},\
  and\ \citenamefont {Tartakovskii}}]{Alexeev2019}%
  \BibitemOpen
  \bibfield  {author} {\bibinfo {author} {\bibfnamefont {E.~M.}\ \bibnamefont
  {Alexeev}}, \bibinfo {author} {\bibfnamefont {D.~A.}\ \bibnamefont
  {Ruiz-Tijerina}}, \bibinfo {author} {\bibfnamefont {M.}~\bibnamefont
  {Danovich}}, \bibinfo {author} {\bibfnamefont {M.~J.}\ \bibnamefont {Hamer}},
  \bibinfo {author} {\bibfnamefont {D.~J.}\ \bibnamefont {Terry}}, \bibinfo
  {author} {\bibfnamefont {P.~K.}\ \bibnamefont {Nayak}}, \bibinfo {author}
  {\bibfnamefont {S.}~\bibnamefont {Ahn}}, \bibinfo {author} {\bibfnamefont
  {S.}~\bibnamefont {Pak}}, \bibinfo {author} {\bibfnamefont {J.}~\bibnamefont
  {Lee}}, \bibinfo {author} {\bibfnamefont {J.~I.}\ \bibnamefont {Sohn}},
  \bibinfo {author} {\bibfnamefont {M.~R.}\ \bibnamefont {Molas}}, \bibinfo
  {author} {\bibfnamefont {M.}~\bibnamefont {Koperski}}, \bibinfo {author}
  {\bibfnamefont {K.}~\bibnamefont {Watanabe}}, \bibinfo {author}
  {\bibfnamefont {T.}~\bibnamefont {Taniguchi}}, \bibinfo {author}
  {\bibfnamefont {K.~S.}\ \bibnamefont {Novoselov}}, \bibinfo {author}
  {\bibfnamefont {R.~V.}\ \bibnamefont {Gorbachev}}, \bibinfo {author}
  {\bibfnamefont {H.~S.}\ \bibnamefont {Shin}}, \bibinfo {author}
  {\bibfnamefont {V.~I.}\ \bibnamefont {Fal'ko}}, \ and\ \bibinfo {author}
  {\bibfnamefont {A.~I.}\ \bibnamefont {Tartakovskii}},\ }\href {\doibase
  10.1038/s41586-019-0986-9} {\bibfield  {journal} {\bibinfo  {journal}
  {Nature}\ }\textbf {\bibinfo {volume} {567}},\ \bibinfo {pages} {81}
  (\bibinfo {year} {2019})}\BibitemShut {NoStop}%
\bibitem [{\citenamefont {Ruiz-Tijerina}\ and\ \citenamefont
  {Fal'ko}(2019)}]{Ruiz2019}%
  \BibitemOpen
  \bibfield  {author} {\bibinfo {author} {\bibfnamefont {D.~A.}\ \bibnamefont
  {Ruiz-Tijerina}}\ and\ \bibinfo {author} {\bibfnamefont {V.~I.}\ \bibnamefont
  {Fal'ko}},\ }\href {\doibase 10.1103/PhysRevB.99.125424} {\bibfield
  {journal} {\bibinfo  {journal} {Phys. Rev. B}\ }\textbf {\bibinfo {volume}
  {99}},\ \bibinfo {pages} {125424} (\bibinfo {year} {2019})}\BibitemShut
  {NoStop}%
\bibitem [{\citenamefont {Shimazaki}\ \emph {et~al.}(2020)\citenamefont
  {Shimazaki}, \citenamefont {Schwartz}, \citenamefont {Watanabe},
  \citenamefont {Taniguchi}, \citenamefont {Kroner},\ and\ \citenamefont
  {Imamo{\u g}lu}}]{Shimazaki2020}%
  \BibitemOpen
  \bibfield  {author} {\bibinfo {author} {\bibfnamefont {Y.}~\bibnamefont
  {Shimazaki}}, \bibinfo {author} {\bibfnamefont {I.}~\bibnamefont {Schwartz}},
  \bibinfo {author} {\bibfnamefont {K.}~\bibnamefont {Watanabe}}, \bibinfo
  {author} {\bibfnamefont {T.}~\bibnamefont {Taniguchi}}, \bibinfo {author}
  {\bibfnamefont {M.}~\bibnamefont {Kroner}}, \ and\ \bibinfo {author}
  {\bibfnamefont {A.}~\bibnamefont {Imamo{\u g}lu}},\ }\href {\doibase
  10.1038/s41586-020-2191-2} {\bibfield  {journal} {\bibinfo  {journal}
  {Nature}\ }\textbf {\bibinfo {volume} {580}},\ \bibinfo {pages} {472}
  (\bibinfo {year} {2020})}\BibitemShut {NoStop}%
\bibitem [{\citenamefont {Carusotto}\ and\ \citenamefont
  {Ciuti}(2013)}]{Carusotto2013}%
  \BibitemOpen
  \bibfield  {author} {\bibinfo {author} {\bibfnamefont {I.}~\bibnamefont
  {Carusotto}}\ and\ \bibinfo {author} {\bibfnamefont {C.}~\bibnamefont
  {Ciuti}},\ }\href {\doibase 10.1103/RevModPhys.85.299} {\bibfield  {journal}
  {\bibinfo  {journal} {Rev. Mod. Phys.}\ }\textbf {\bibinfo {volume} {85}},\
  \bibinfo {pages} {299} (\bibinfo {year} {2013})}\BibitemShut {NoStop}%
\bibitem [{\citenamefont {Brune}\ \emph {et~al.}(1987)\citenamefont {Brune},
  \citenamefont {Raimond},\ and\ \citenamefont {Haroche}}]{Brune1987}%
  \BibitemOpen
  \bibfield  {author} {\bibinfo {author} {\bibfnamefont {M.}~\bibnamefont
  {Brune}}, \bibinfo {author} {\bibfnamefont {J.~M.}\ \bibnamefont {Raimond}},
  \ and\ \bibinfo {author} {\bibfnamefont {S.}~\bibnamefont {Haroche}},\ }\href
  {\doibase 10.1103/PhysRevA.35.154} {\bibfield  {journal} {\bibinfo  {journal}
  {Phys. Rev. A}\ }\textbf {\bibinfo {volume} {35}},\ \bibinfo {pages} {154}
  (\bibinfo {year} {1987})}\BibitemShut {NoStop}%
\bibitem [{\citenamefont {Davidovich}\ \emph {et~al.}(1987)\citenamefont
  {Davidovich}, \citenamefont {Raimond}, \citenamefont {Brune},\ and\
  \citenamefont {Haroche}}]{Davidovich1987}%
  \BibitemOpen
  \bibfield  {author} {\bibinfo {author} {\bibfnamefont {L.}~\bibnamefont
  {Davidovich}}, \bibinfo {author} {\bibfnamefont {J.~M.}\ \bibnamefont
  {Raimond}}, \bibinfo {author} {\bibfnamefont {M.}~\bibnamefont {Brune}}, \
  and\ \bibinfo {author} {\bibfnamefont {S.}~\bibnamefont {Haroche}},\ }\href
  {\doibase 10.1103/PhysRevA.36.3771} {\bibfield  {journal} {\bibinfo
  {journal} {Phys. Rev. A}\ }\textbf {\bibinfo {volume} {36}},\ \bibinfo
  {pages} {3771} (\bibinfo {year} {1987})}\BibitemShut {NoStop}%
\bibitem [{\citenamefont {Lewenstein}\ \emph {et~al.}(1990)\citenamefont
  {Lewenstein}, \citenamefont {Zhu},\ and\ \citenamefont
  {Mossberg}}]{Lewenstein1990}%
  \BibitemOpen
  \bibfield  {author} {\bibinfo {author} {\bibfnamefont {M.}~\bibnamefont
  {Lewenstein}}, \bibinfo {author} {\bibfnamefont {Y.}~\bibnamefont {Zhu}}, \
  and\ \bibinfo {author} {\bibfnamefont {T.~W.}\ \bibnamefont {Mossberg}},\
  }\href {\doibase 10.1103/PhysRevLett.64.3131} {\bibfield  {journal} {\bibinfo
   {journal} {Phys. Rev. Lett.}\ }\textbf {\bibinfo {volume} {64}},\ \bibinfo
  {pages} {3131} (\bibinfo {year} {1990})}\BibitemShut {NoStop}%
\bibitem [{\citenamefont {Orszag}\ \emph {et~al.}(1993)\citenamefont {Orszag},
  \citenamefont {Roa},\ and\ \citenamefont {Ram\'{\i}rez}}]{Orszag1993}%
  \BibitemOpen
  \bibfield  {author} {\bibinfo {author} {\bibfnamefont {M.}~\bibnamefont
  {Orszag}}, \bibinfo {author} {\bibfnamefont {L.}~\bibnamefont {Roa}}, \ and\
  \bibinfo {author} {\bibfnamefont {R.}~\bibnamefont {Ram\'{\i}rez}},\ }\href
  {\doibase 10.1103/PhysRevA.48.4648} {\bibfield  {journal} {\bibinfo
  {journal} {Phys. Rev. A}\ }\textbf {\bibinfo {volume} {48}},\ \bibinfo
  {pages} {4648} (\bibinfo {year} {1993})}\BibitemShut {NoStop}%
\bibitem [{\citenamefont {Kasprzak}\ \emph {et~al.}(2006)\citenamefont
  {Kasprzak}, \citenamefont {Richard}, \citenamefont {Kundermann},
  \citenamefont {Baas}, \citenamefont {Jeambrun}, \citenamefont {Keeling},
  \citenamefont {Marchetti}, \citenamefont {Szyma{\'n}ska}, \citenamefont
  {Andr{\'e}}, \citenamefont {Staehli}, \citenamefont {Savona}, \citenamefont
  {Littlewood}, \citenamefont {Deveaud},\ and\ \citenamefont
  {Dang}}]{Kasprzak2006}%
  \BibitemOpen
  \bibfield  {author} {\bibinfo {author} {\bibfnamefont {J.}~\bibnamefont
  {Kasprzak}}, \bibinfo {author} {\bibfnamefont {M.}~\bibnamefont {Richard}},
  \bibinfo {author} {\bibfnamefont {S.}~\bibnamefont {Kundermann}}, \bibinfo
  {author} {\bibfnamefont {A.}~\bibnamefont {Baas}}, \bibinfo {author}
  {\bibfnamefont {P.}~\bibnamefont {Jeambrun}}, \bibinfo {author}
  {\bibfnamefont {J.~M.~J.}\ \bibnamefont {Keeling}}, \bibinfo {author}
  {\bibfnamefont {F.~M.}\ \bibnamefont {Marchetti}}, \bibinfo {author}
  {\bibfnamefont {M.~H.}\ \bibnamefont {Szyma{\'n}ska}}, \bibinfo {author}
  {\bibfnamefont {R.}~\bibnamefont {Andr{\'e}}}, \bibinfo {author}
  {\bibfnamefont {J.~L.}\ \bibnamefont {Staehli}}, \bibinfo {author}
  {\bibfnamefont {V.}~\bibnamefont {Savona}}, \bibinfo {author} {\bibfnamefont
  {P.~B.}\ \bibnamefont {Littlewood}}, \bibinfo {author} {\bibfnamefont
  {B.}~\bibnamefont {Deveaud}}, \ and\ \bibinfo {author} {\bibfnamefont
  {L.~S.}\ \bibnamefont {Dang}},\ }\href {\doibase 10.1038/nature05131}
  {\bibfield  {journal} {\bibinfo  {journal} {Nature}\ }\textbf {\bibinfo
  {volume} {443}},\ \bibinfo {pages} {409} (\bibinfo {year}
  {2006})}\BibitemShut {NoStop}%
\bibitem [{\citenamefont {Deng}\ \emph {et~al.}(2010)\citenamefont {Deng},
  \citenamefont {Haug},\ and\ \citenamefont {Yamamoto}}]{Deng2010}%
  \BibitemOpen
  \bibfield  {author} {\bibinfo {author} {\bibfnamefont {H.}~\bibnamefont
  {Deng}}, \bibinfo {author} {\bibfnamefont {H.}~\bibnamefont {Haug}}, \ and\
  \bibinfo {author} {\bibfnamefont {Y.}~\bibnamefont {Yamamoto}},\ }\href
  {\doibase 10.1103/RevModPhys.82.1489} {\bibfield  {journal} {\bibinfo
  {journal} {Rev. Mod. Phys.}\ }\textbf {\bibinfo {volume} {82}},\ \bibinfo
  {pages} {1489} (\bibinfo {year} {2010})}\BibitemShut {NoStop}%
\bibitem [{\citenamefont {Amo}\ \emph {et~al.}(2009)\citenamefont {Amo},
  \citenamefont {Lefr{\`e}re}, \citenamefont {Pigeon}, \citenamefont {Adrados},
  \citenamefont {Ciuti}, \citenamefont {Carusotto}, \citenamefont {Houdr{\'e}},
  \citenamefont {Giacobino},\ and\ \citenamefont {Bramati}}]{Amo2009}%
  \BibitemOpen
  \bibfield  {author} {\bibinfo {author} {\bibfnamefont {A.}~\bibnamefont
  {Amo}}, \bibinfo {author} {\bibfnamefont {J.}~\bibnamefont {Lefr{\`e}re}},
  \bibinfo {author} {\bibfnamefont {S.}~\bibnamefont {Pigeon}}, \bibinfo
  {author} {\bibfnamefont {C.}~\bibnamefont {Adrados}}, \bibinfo {author}
  {\bibfnamefont {C.}~\bibnamefont {Ciuti}}, \bibinfo {author} {\bibfnamefont
  {I.}~\bibnamefont {Carusotto}}, \bibinfo {author} {\bibfnamefont
  {R.}~\bibnamefont {Houdr{\'e}}}, \bibinfo {author} {\bibfnamefont
  {E.}~\bibnamefont {Giacobino}}, \ and\ \bibinfo {author} {\bibfnamefont
  {A.}~\bibnamefont {Bramati}},\ }\href {\doibase 10.1038/nphys1364} {\bibfield
   {journal} {\bibinfo  {journal} {Nature Physics}\ }\textbf {\bibinfo {volume}
  {5}},\ \bibinfo {pages} {805} (\bibinfo {year} {2009})}\BibitemShut {NoStop}%
\bibitem [{\citenamefont {Dominici}\ \emph {et~al.}(2018)\citenamefont
  {Dominici}, \citenamefont {Carretero-Gonz{\'a}lez}, \citenamefont
  {Gianfrate}, \citenamefont {Cuevas-Maraver}, \citenamefont {Rodrigues},
  \citenamefont {Frantzeskakis}, \citenamefont {Lerario}, \citenamefont
  {Ballarini}, \citenamefont {De~Giorgi}, \citenamefont {Gigli}, \citenamefont
  {Kevrekidis},\ and\ \citenamefont {Sanvitto}}]{Dominici2018}%
  \BibitemOpen
  \bibfield  {author} {\bibinfo {author} {\bibfnamefont {L.}~\bibnamefont
  {Dominici}}, \bibinfo {author} {\bibfnamefont {R.}~\bibnamefont
  {Carretero-Gonz{\'a}lez}}, \bibinfo {author} {\bibfnamefont {A.}~\bibnamefont
  {Gianfrate}}, \bibinfo {author} {\bibfnamefont {J.}~\bibnamefont
  {Cuevas-Maraver}}, \bibinfo {author} {\bibfnamefont {A.~S.}\ \bibnamefont
  {Rodrigues}}, \bibinfo {author} {\bibfnamefont {D.~J.}\ \bibnamefont
  {Frantzeskakis}}, \bibinfo {author} {\bibfnamefont {G.}~\bibnamefont
  {Lerario}}, \bibinfo {author} {\bibfnamefont {D.}~\bibnamefont {Ballarini}},
  \bibinfo {author} {\bibfnamefont {M.}~\bibnamefont {De~Giorgi}}, \bibinfo
  {author} {\bibfnamefont {G.}~\bibnamefont {Gigli}}, \bibinfo {author}
  {\bibfnamefont {P.~G.}\ \bibnamefont {Kevrekidis}}, \ and\ \bibinfo {author}
  {\bibfnamefont {D.}~\bibnamefont {Sanvitto}},\ }\href {\doibase
  10.1038/s41467-018-03736-5} {\bibfield  {journal} {\bibinfo  {journal}
  {Nature Communications}\ }\textbf {\bibinfo {volume} {9}},\ \bibinfo {pages}
  {1467} (\bibinfo {year} {2018})}\BibitemShut {NoStop}%
\bibitem [{\citenamefont {Takemura}\ \emph {et~al.}(2014)\citenamefont
  {Takemura}, \citenamefont {Trebaol}, \citenamefont {Wouters}, \citenamefont
  {Portella-Oberli},\ and\ \citenamefont {Deveaud}}]{Takemura2014}%
  \BibitemOpen
  \bibfield  {author} {\bibinfo {author} {\bibfnamefont {N.}~\bibnamefont
  {Takemura}}, \bibinfo {author} {\bibfnamefont {S.}~\bibnamefont {Trebaol}},
  \bibinfo {author} {\bibfnamefont {M.}~\bibnamefont {Wouters}}, \bibinfo
  {author} {\bibfnamefont {M.~T.}\ \bibnamefont {Portella-Oberli}}, \ and\
  \bibinfo {author} {\bibfnamefont {B.}~\bibnamefont {Deveaud}},\ }\href
  {\doibase 10.1038/nphys2999} {\bibfield  {journal} {\bibinfo  {journal}
  {Nature Physics}\ }\textbf {\bibinfo {volume} {10}},\ \bibinfo {pages} {500}
  (\bibinfo {year} {2014})}\BibitemShut {NoStop}%
\bibitem [{\citenamefont {Sidler}\ \emph {et~al.}(2017)\citenamefont {Sidler},
  \citenamefont {Back}, \citenamefont {Cotlet}, \citenamefont {Srivastava},
  \citenamefont {Fink}, \citenamefont {Kroner}, \citenamefont {Demler},\ and\
  \citenamefont {Imamoglu}}]{Sidler2017}%
  \BibitemOpen
  \bibfield  {author} {\bibinfo {author} {\bibfnamefont {M.}~\bibnamefont
  {Sidler}}, \bibinfo {author} {\bibfnamefont {P.}~\bibnamefont {Back}},
  \bibinfo {author} {\bibfnamefont {O.}~\bibnamefont {Cotlet}}, \bibinfo
  {author} {\bibfnamefont {A.}~\bibnamefont {Srivastava}}, \bibinfo {author}
  {\bibfnamefont {T.}~\bibnamefont {Fink}}, \bibinfo {author} {\bibfnamefont
  {M.}~\bibnamefont {Kroner}}, \bibinfo {author} {\bibfnamefont
  {E.}~\bibnamefont {Demler}}, \ and\ \bibinfo {author} {\bibfnamefont
  {A.}~\bibnamefont {Imamoglu}},\ }\href {\doibase 10.1038/nphys3949}
  {\bibfield  {journal} {\bibinfo  {journal} {Nature Physics}\ }\textbf
  {\bibinfo {volume} {13}},\ \bibinfo {pages} {255} (\bibinfo {year}
  {2017})}\BibitemShut {NoStop}%
\bibitem [{\citenamefont {Emmanuele}\ \emph {et~al.}(2020)\citenamefont
  {Emmanuele}, \citenamefont {Sich}, \citenamefont {Kyriienko}, \citenamefont
  {Shahnazaryan}, \citenamefont {Withers}, \citenamefont {Catanzaro},
  \citenamefont {Walker}, \citenamefont {Benimetskiy}, \citenamefont
  {Skolnick}, \citenamefont {Tartakovskii}, \citenamefont {Shelykh},\ and\
  \citenamefont {Krizhanovskii}}]{Emmanuele2020}%
  \BibitemOpen
  \bibfield  {author} {\bibinfo {author} {\bibfnamefont {R.~P.~A.}\
  \bibnamefont {Emmanuele}}, \bibinfo {author} {\bibfnamefont {M.}~\bibnamefont
  {Sich}}, \bibinfo {author} {\bibfnamefont {O.}~\bibnamefont {Kyriienko}},
  \bibinfo {author} {\bibfnamefont {V.}~\bibnamefont {Shahnazaryan}}, \bibinfo
  {author} {\bibfnamefont {F.}~\bibnamefont {Withers}}, \bibinfo {author}
  {\bibfnamefont {A.}~\bibnamefont {Catanzaro}}, \bibinfo {author}
  {\bibfnamefont {P.~M.}\ \bibnamefont {Walker}}, \bibinfo {author}
  {\bibfnamefont {F.~A.}\ \bibnamefont {Benimetskiy}}, \bibinfo {author}
  {\bibfnamefont {M.~S.}\ \bibnamefont {Skolnick}}, \bibinfo {author}
  {\bibfnamefont {A.~I.}\ \bibnamefont {Tartakovskii}}, \bibinfo {author}
  {\bibfnamefont {I.~A.}\ \bibnamefont {Shelykh}}, \ and\ \bibinfo {author}
  {\bibfnamefont {D.~N.}\ \bibnamefont {Krizhanovskii}},\ }\href {\doibase
  10.1038/s41467-020-17340-z} {\bibfield  {journal} {\bibinfo  {journal}
  {Nature Communications}\ }\textbf {\bibinfo {volume} {11}},\ \bibinfo {pages}
  {3589} (\bibinfo {year} {2020})}\BibitemShut {NoStop}%
\bibitem [{\citenamefont {Tan}\ \emph {et~al.}(2020)\citenamefont {Tan},
  \citenamefont {Cotlet}, \citenamefont {Bergschneider}, \citenamefont
  {Schmidt}, \citenamefont {Back}, \citenamefont {Shimazaki}, \citenamefont
  {Kroner},\ and\ \citenamefont {\ifmmode \dot{I}\else
  \.{I}\fi{}mamo\ifmmode~\breve{g}\else \u{g}\fi{}lu}}]{Tan2020}%
  \BibitemOpen
  \bibfield  {author} {\bibinfo {author} {\bibfnamefont {L.~B.}\ \bibnamefont
  {Tan}}, \bibinfo {author} {\bibfnamefont {O.}~\bibnamefont {Cotlet}},
  \bibinfo {author} {\bibfnamefont {A.}~\bibnamefont {Bergschneider}}, \bibinfo
  {author} {\bibfnamefont {R.}~\bibnamefont {Schmidt}}, \bibinfo {author}
  {\bibfnamefont {P.}~\bibnamefont {Back}}, \bibinfo {author} {\bibfnamefont
  {Y.}~\bibnamefont {Shimazaki}}, \bibinfo {author} {\bibfnamefont
  {M.}~\bibnamefont {Kroner}}, \ and\ \bibinfo {author} {\bibfnamefont
  {A.~m.~c.}\ \bibnamefont {\ifmmode \dot{I}\else
  \.{I}\fi{}mamo\ifmmode~\breve{g}\else \u{g}\fi{}lu}},\ }\href {\doibase
  10.1103/PhysRevX.10.021011} {\bibfield  {journal} {\bibinfo  {journal} {Phys.
  Rev. X}\ }\textbf {\bibinfo {volume} {10}},\ \bibinfo {pages} {021011}
  (\bibinfo {year} {2020})}\BibitemShut {NoStop}%
\bibitem [{\citenamefont {Camacho-Guardian}\ \emph {et~al.}(2021)\citenamefont
  {Camacho-Guardian}, \citenamefont {Bastarrachea-Magnani},\ and\ \citenamefont
  {Bruun}}]{Camacho2021}%
  \BibitemOpen
  \bibfield  {author} {\bibinfo {author} {\bibfnamefont {A.}~\bibnamefont
  {Camacho-Guardian}}, \bibinfo {author} {\bibfnamefont {M.~A.}\ \bibnamefont
  {Bastarrachea-Magnani}}, \ and\ \bibinfo {author} {\bibfnamefont {G.~M.}\
  \bibnamefont {Bruun}},\ }\href {\doibase 10.1103/PhysRevLett.126.017401}
  {\bibfield  {journal} {\bibinfo  {journal} {Phys. Rev. Lett.}\ }\textbf
  {\bibinfo {volume} {126}},\ \bibinfo {pages} {017401} (\bibinfo {year}
  {2021})}\BibitemShut {NoStop}%
\bibitem [{\citenamefont {Zhang}\ \emph {et~al.}(2021)\citenamefont {Zhang},
  \citenamefont {Wu}, \citenamefont {Hou}, \citenamefont {Zhang}, \citenamefont
  {Chou}, \citenamefont {Watanabe}, \citenamefont {Taniguchi}, \citenamefont
  {Forrest},\ and\ \citenamefont {Deng}}]{Zhang2021}%
  \BibitemOpen
  \bibfield  {author} {\bibinfo {author} {\bibfnamefont {L.}~\bibnamefont
  {Zhang}}, \bibinfo {author} {\bibfnamefont {F.}~\bibnamefont {Wu}}, \bibinfo
  {author} {\bibfnamefont {S.}~\bibnamefont {Hou}}, \bibinfo {author}
  {\bibfnamefont {Z.}~\bibnamefont {Zhang}}, \bibinfo {author} {\bibfnamefont
  {Y.-H.}\ \bibnamefont {Chou}}, \bibinfo {author} {\bibfnamefont
  {K.}~\bibnamefont {Watanabe}}, \bibinfo {author} {\bibfnamefont
  {T.}~\bibnamefont {Taniguchi}}, \bibinfo {author} {\bibfnamefont {S.~R.}\
  \bibnamefont {Forrest}}, \ and\ \bibinfo {author} {\bibfnamefont
  {H.}~\bibnamefont {Deng}},\ }\href {\doibase 10.1038/s41586-021-03228-5}
  {\bibfield  {journal} {\bibinfo  {journal} {Nature}\ }\textbf {\bibinfo
  {volume} {591}},\ \bibinfo {pages} {61} (\bibinfo {year} {2021})}\BibitemShut
  {NoStop}%
\bibitem [{SM()}]{SM}%
  \BibitemOpen
  \href@noop {} {}\bibinfo {note} {See {\it Supplemental Material} online for
  details.}\BibitemShut {Stop}%
\bibitem [{\citenamefont {Brennecke}\ \emph {et~al.}(2007)\citenamefont
  {Brennecke}, \citenamefont {Donner}, \citenamefont {Ritter}, \citenamefont
  {Bourdel}, \citenamefont {K{\"o}hl},\ and\ \citenamefont
  {Esslinger}}]{Brennecke2007}%
  \BibitemOpen
  \bibfield  {author} {\bibinfo {author} {\bibfnamefont {F.}~\bibnamefont
  {Brennecke}}, \bibinfo {author} {\bibfnamefont {T.}~\bibnamefont {Donner}},
  \bibinfo {author} {\bibfnamefont {S.}~\bibnamefont {Ritter}}, \bibinfo
  {author} {\bibfnamefont {T.}~\bibnamefont {Bourdel}}, \bibinfo {author}
  {\bibfnamefont {M.}~\bibnamefont {K{\"o}hl}}, \ and\ \bibinfo {author}
  {\bibfnamefont {T.}~\bibnamefont {Esslinger}},\ }\href {\doibase
  10.1038/nature06120} {\bibfield  {journal} {\bibinfo  {journal} {Nature}\
  }\textbf {\bibinfo {volume} {450}},\ \bibinfo {pages} {268} (\bibinfo {year}
  {2007})}\BibitemShut {NoStop}%
\bibitem [{\citenamefont {Ritsch}\ \emph {et~al.}(2013)\citenamefont {Ritsch},
  \citenamefont {Domokos}, \citenamefont {Brennecke},\ and\ \citenamefont
  {Esslinger}}]{Ritsch2013}%
  \BibitemOpen
  \bibfield  {author} {\bibinfo {author} {\bibfnamefont {H.}~\bibnamefont
  {Ritsch}}, \bibinfo {author} {\bibfnamefont {P.}~\bibnamefont {Domokos}},
  \bibinfo {author} {\bibfnamefont {F.}~\bibnamefont {Brennecke}}, \ and\
  \bibinfo {author} {\bibfnamefont {T.}~\bibnamefont {Esslinger}},\ }\href
  {\doibase 10.1103/RevModPhys.85.553} {\bibfield  {journal} {\bibinfo
  {journal} {Rev. Mod. Phys.}\ }\textbf {\bibinfo {volume} {85}},\ \bibinfo
  {pages} {553} (\bibinfo {year} {2013})}\BibitemShut {NoStop}%
\bibitem [{\citenamefont {{Mekhov}}\ and\ \citenamefont
  {{Ritsch}}(2012)}]{Mekhov2012}%
  \BibitemOpen
  \bibfield  {author} {\bibinfo {author} {\bibfnamefont {I.~B.}\ \bibnamefont
  {{Mekhov}}}\ and\ \bibinfo {author} {\bibfnamefont {H.}~\bibnamefont
  {{Ritsch}}},\ }\href {\doibase 10.1088/0953-4075/45/10/102001} {\bibfield
  {journal} {\bibinfo  {journal} {Journal of Physics B Atomic Molecular
  Physics}\ }\textbf {\bibinfo {volume} {45}},\ \bibinfo {eid} {102001}
  (\bibinfo {year} {2012})}\BibitemShut {NoStop}%
\bibitem [{\citenamefont {Baumann}\ \emph {et~al.}(2010)\citenamefont
  {Baumann}, \citenamefont {Guerlin}, \citenamefont {Brennecke},\ and\
  \citenamefont {Esslinger}}]{Baumann2010}%
  \BibitemOpen
  \bibfield  {author} {\bibinfo {author} {\bibfnamefont {K.}~\bibnamefont
  {Baumann}}, \bibinfo {author} {\bibfnamefont {C.}~\bibnamefont {Guerlin}},
  \bibinfo {author} {\bibfnamefont {F.}~\bibnamefont {Brennecke}}, \ and\
  \bibinfo {author} {\bibfnamefont {T.}~\bibnamefont {Esslinger}},\ }\href
  {\doibase 10.1038/nature09009} {\bibfield  {journal} {\bibinfo  {journal}
  {Nature}\ }\textbf {\bibinfo {volume} {464}},\ \bibinfo {pages} {1301}
  (\bibinfo {year} {2010})}\BibitemShut {NoStop}%
\bibitem [{\citenamefont {L{\'e}onard}\ \emph {et~al.}(2017)\citenamefont
  {L{\'e}onard}, \citenamefont {Morales}, \citenamefont {Zupancic},
  \citenamefont {Esslinger},\ and\ \citenamefont {Donner}}]{Leonard2017}%
  \BibitemOpen
  \bibfield  {author} {\bibinfo {author} {\bibfnamefont {J.}~\bibnamefont
  {L{\'e}onard}}, \bibinfo {author} {\bibfnamefont {A.}~\bibnamefont
  {Morales}}, \bibinfo {author} {\bibfnamefont {P.}~\bibnamefont {Zupancic}},
  \bibinfo {author} {\bibfnamefont {T.}~\bibnamefont {Esslinger}}, \ and\
  \bibinfo {author} {\bibfnamefont {T.}~\bibnamefont {Donner}},\ }\href
  {\doibase 10.1038/nature21067} {\bibfield  {journal} {\bibinfo  {journal}
  {Nature}\ }\textbf {\bibinfo {volume} {543}},\ \bibinfo {pages} {87}
  (\bibinfo {year} {2017})}\BibitemShut {NoStop}%
\bibitem [{\citenamefont {Eastham}\ and\ \citenamefont
  {Littlewood}(2001)}]{Eastham2001}%
  \BibitemOpen
  \bibfield  {author} {\bibinfo {author} {\bibfnamefont {P.~R.}\ \bibnamefont
  {Eastham}}\ and\ \bibinfo {author} {\bibfnamefont {P.~B.}\ \bibnamefont
  {Littlewood}},\ }\href {\doibase 10.1103/PhysRevB.64.235101} {\bibfield
  {journal} {\bibinfo  {journal} {Phys. Rev. B}\ }\textbf {\bibinfo {volume}
  {64}},\ \bibinfo {pages} {235101} (\bibinfo {year} {2001})}\BibitemShut
  {NoStop}%
\bibitem [{\citenamefont {Szymanska}\ \emph {et~al.}(2003)\citenamefont
  {Szymanska}, \citenamefont {Littlewood},\ and\ \citenamefont
  {Simons}}]{Szymanska2003}%
  \BibitemOpen
  \bibfield  {author} {\bibinfo {author} {\bibfnamefont {M.~H.}\ \bibnamefont
  {Szymanska}}, \bibinfo {author} {\bibfnamefont {P.~B.}\ \bibnamefont
  {Littlewood}}, \ and\ \bibinfo {author} {\bibfnamefont {B.~D.}\ \bibnamefont
  {Simons}},\ }\href {\doibase 10.1103/PhysRevA.68.013818} {\bibfield
  {journal} {\bibinfo  {journal} {Phys. Rev. A}\ }\textbf {\bibinfo {volume}
  {68}},\ \bibinfo {pages} {013818} (\bibinfo {year} {2003})}\BibitemShut
  {NoStop}%
\bibitem [{\citenamefont {Keeling}\ \emph {et~al.}(2004)\citenamefont
  {Keeling}, \citenamefont {Eastham}, \citenamefont {Szymanska},\ and\
  \citenamefont {Littlewood}}]{Keeling2004}%
  \BibitemOpen
  \bibfield  {author} {\bibinfo {author} {\bibfnamefont {J.}~\bibnamefont
  {Keeling}}, \bibinfo {author} {\bibfnamefont {P.~R.}\ \bibnamefont
  {Eastham}}, \bibinfo {author} {\bibfnamefont {M.~H.}\ \bibnamefont
  {Szymanska}}, \ and\ \bibinfo {author} {\bibfnamefont {P.~B.}\ \bibnamefont
  {Littlewood}},\ }\href {\doibase 10.1103/PhysRevLett.93.226403} {\bibfield
  {journal} {\bibinfo  {journal} {Phys. Rev. Lett.}\ }\textbf {\bibinfo
  {volume} {93}},\ \bibinfo {pages} {226403} (\bibinfo {year}
  {2004})}\BibitemShut {NoStop}%
\bibitem [{\citenamefont {Drummond}\ and\ \citenamefont
  {Walls}(1980)}]{Drummond1980}%
  \BibitemOpen
  \bibfield  {author} {\bibinfo {author} {\bibfnamefont {P.~D.}\ \bibnamefont
  {Drummond}}\ and\ \bibinfo {author} {\bibfnamefont {D.~F.}\ \bibnamefont
  {Walls}},\ }\href {\doibase 10.1088/0305-4470/13/2/034} {\bibfield  {journal}
  {\bibinfo  {journal} {Journal of Physics A: Mathematical and General}\
  }\textbf {\bibinfo {volume} {13}},\ \bibinfo {pages} {725} (\bibinfo {year}
  {1980})}\BibitemShut {NoStop}%
\bibitem [{\citenamefont {Le~Boit\'e}\ \emph {et~al.}(2013)\citenamefont
  {Le~Boit\'e}, \citenamefont {Orso},\ and\ \citenamefont {Ciuti}}]{Boite2013}%
  \BibitemOpen
  \bibfield  {author} {\bibinfo {author} {\bibfnamefont {A.}~\bibnamefont
  {Le~Boit\'e}}, \bibinfo {author} {\bibfnamefont {G.}~\bibnamefont {Orso}}, \
  and\ \bibinfo {author} {\bibfnamefont {C.}~\bibnamefont {Ciuti}},\ }\href
  {\doibase 10.1103/PhysRevLett.110.233601} {\bibfield  {journal} {\bibinfo
  {journal} {Phys. Rev. Lett.}\ }\textbf {\bibinfo {volume} {110}},\ \bibinfo
  {pages} {233601} (\bibinfo {year} {2013})}\BibitemShut {NoStop}%
\bibitem [{\citenamefont {Le~Boit\'e}\ \emph {et~al.}(2014)\citenamefont
  {Le~Boit\'e}, \citenamefont {Orso},\ and\ \citenamefont {Ciuti}}]{Boite2014}%
  \BibitemOpen
  \bibfield  {author} {\bibinfo {author} {\bibfnamefont {A.}~\bibnamefont
  {Le~Boit\'e}}, \bibinfo {author} {\bibfnamefont {G.}~\bibnamefont {Orso}}, \
  and\ \bibinfo {author} {\bibfnamefont {C.}~\bibnamefont {Ciuti}},\ }\href
  {\doibase 10.1103/PhysRevA.90.063821} {\bibfield  {journal} {\bibinfo
  {journal} {Phys. Rev. A}\ }\textbf {\bibinfo {volume} {90}},\ \bibinfo
  {pages} {063821} (\bibinfo {year} {2014})}\BibitemShut {NoStop}%
\bibitem [{\citenamefont {Biondi}\ \emph {et~al.}(2017)\citenamefont {Biondi},
  \citenamefont {Blatter}, \citenamefont {T\"ureci},\ and\ \citenamefont
  {Schmidt}}]{Biondi2017}%
  \BibitemOpen
  \bibfield  {author} {\bibinfo {author} {\bibfnamefont {M.}~\bibnamefont
  {Biondi}}, \bibinfo {author} {\bibfnamefont {G.}~\bibnamefont {Blatter}},
  \bibinfo {author} {\bibfnamefont {H.~E.}\ \bibnamefont {T\"ureci}}, \ and\
  \bibinfo {author} {\bibfnamefont {S.}~\bibnamefont {Schmidt}},\ }\href
  {\doibase 10.1103/PhysRevA.96.043809} {\bibfield  {journal} {\bibinfo
  {journal} {Phys. Rev. A}\ }\textbf {\bibinfo {volume} {96}},\ \bibinfo
  {pages} {043809} (\bibinfo {year} {2017})}\BibitemShut {NoStop}%
\bibitem [{\citenamefont {Ciuti}\ and\ \citenamefont
  {Carusotto}(2005)}]{Ciuti2005}%
  \BibitemOpen
  \bibfield  {author} {\bibinfo {author} {\bibfnamefont {C.}~\bibnamefont
  {Ciuti}}\ and\ \bibinfo {author} {\bibfnamefont {I.}~\bibnamefont
  {Carusotto}},\ }\href {\doibase https://doi.org/10.1002/pssb.200560961}
  {\bibfield  {journal} {\bibinfo  {journal} {Physica Status Solidi (b)}\
  }\textbf {\bibinfo {volume} {242}},\ \bibinfo {pages} {2224} (\bibinfo {year}
  {2005})}\BibitemShut {NoStop}%
\bibitem [{\citenamefont {Wen}\ \emph {et~al.}(2020)\citenamefont {Wen},
  \citenamefont {Wu},\ and\ \citenamefont {Yu}}]{Wen2020}%
  \BibitemOpen
  \bibfield  {author} {\bibinfo {author} {\bibfnamefont {W.}~\bibnamefont
  {Wen}}, \bibinfo {author} {\bibfnamefont {L.}~\bibnamefont {Wu}}, \ and\
  \bibinfo {author} {\bibfnamefont {T.}~\bibnamefont {Yu}},\ }\bibfield
  {booktitle} {\emph {\bibinfo {booktitle} {ACS Materials Letters}},\ }\href
  {\doibase 10.1021/acsmaterialslett.0c00277} {\bibfield  {journal} {\bibinfo
  {journal} {ACS Materials Letters}\ }\textbf {\bibinfo {volume} {2}},\
  \bibinfo {pages} {1328} (\bibinfo {year} {2020})}\BibitemShut {NoStop}%
\bibitem [{\citenamefont {Wu}\ \emph {et~al.}(2015)\citenamefont {Wu},
  \citenamefont {Buckley}, \citenamefont {Schaibley}, \citenamefont {Feng},
  \citenamefont {Yan}, \citenamefont {Mandrus}, \citenamefont {Hatami},
  \citenamefont {Yao}, \citenamefont {Vu{\v c}kovi{\'c}}, \citenamefont
  {Majumdar},\ and\ \citenamefont {Xu}}]{Wu2015}%
  \BibitemOpen
  \bibfield  {author} {\bibinfo {author} {\bibfnamefont {S.}~\bibnamefont
  {Wu}}, \bibinfo {author} {\bibfnamefont {S.}~\bibnamefont {Buckley}},
  \bibinfo {author} {\bibfnamefont {J.~R.}\ \bibnamefont {Schaibley}}, \bibinfo
  {author} {\bibfnamefont {L.}~\bibnamefont {Feng}}, \bibinfo {author}
  {\bibfnamefont {J.}~\bibnamefont {Yan}}, \bibinfo {author} {\bibfnamefont
  {D.~G.}\ \bibnamefont {Mandrus}}, \bibinfo {author} {\bibfnamefont
  {F.}~\bibnamefont {Hatami}}, \bibinfo {author} {\bibfnamefont
  {W.}~\bibnamefont {Yao}}, \bibinfo {author} {\bibfnamefont {J.}~\bibnamefont
  {Vu{\v c}kovi{\'c}}}, \bibinfo {author} {\bibfnamefont {A.}~\bibnamefont
  {Majumdar}}, \ and\ \bibinfo {author} {\bibfnamefont {X.}~\bibnamefont
  {Xu}},\ }\href {\doibase 10.1038/nature14290} {\bibfield  {journal} {\bibinfo
   {journal} {Nature}\ }\textbf {\bibinfo {volume} {520}},\ \bibinfo {pages}
  {69} (\bibinfo {year} {2015})}\BibitemShut {NoStop}%
\bibitem [{\citenamefont {Ye}\ \emph {et~al.}(2015)\citenamefont {Ye},
  \citenamefont {Wong}, \citenamefont {Lu}, \citenamefont {Ni}, \citenamefont
  {Zhu}, \citenamefont {Chen}, \citenamefont {Wang},\ and\ \citenamefont
  {Zhang}}]{Ye2015}%
  \BibitemOpen
  \bibfield  {author} {\bibinfo {author} {\bibfnamefont {Y.}~\bibnamefont
  {Ye}}, \bibinfo {author} {\bibfnamefont {Z.~J.}\ \bibnamefont {Wong}},
  \bibinfo {author} {\bibfnamefont {X.}~\bibnamefont {Lu}}, \bibinfo {author}
  {\bibfnamefont {X.}~\bibnamefont {Ni}}, \bibinfo {author} {\bibfnamefont
  {H.}~\bibnamefont {Zhu}}, \bibinfo {author} {\bibfnamefont {X.}~\bibnamefont
  {Chen}}, \bibinfo {author} {\bibfnamefont {Y.}~\bibnamefont {Wang}}, \ and\
  \bibinfo {author} {\bibfnamefont {X.}~\bibnamefont {Zhang}},\ }\href
  {\doibase 10.1038/nphoton.2015.197} {\bibfield  {journal} {\bibinfo
  {journal} {Nature Photonics}\ }\textbf {\bibinfo {volume} {9}},\ \bibinfo
  {pages} {733} (\bibinfo {year} {2015})}\BibitemShut {NoStop}%
\bibitem [{\citenamefont {Salehzadeh}\ \emph {et~al.}(2015)\citenamefont
  {Salehzadeh}, \citenamefont {Djavid}, \citenamefont {Tran}, \citenamefont
  {Shih},\ and\ \citenamefont {Mi}}]{Salehzadeh2015}%
  \BibitemOpen
  \bibfield  {author} {\bibinfo {author} {\bibfnamefont {O.}~\bibnamefont
  {Salehzadeh}}, \bibinfo {author} {\bibfnamefont {M.}~\bibnamefont {Djavid}},
  \bibinfo {author} {\bibfnamefont {N.~H.}\ \bibnamefont {Tran}}, \bibinfo
  {author} {\bibfnamefont {I.}~\bibnamefont {Shih}}, \ and\ \bibinfo {author}
  {\bibfnamefont {Z.}~\bibnamefont {Mi}},\ }\bibfield  {booktitle} {\emph
  {\bibinfo {booktitle} {Nano Letters}},\ }\href {\doibase
  10.1021/acs.nanolett.5b01665} {\bibfield  {journal} {\bibinfo  {journal}
  {Nano Letters}\ }\textbf {\bibinfo {volume} {15}},\ \bibinfo {pages} {5302}
  (\bibinfo {year} {2015})}\BibitemShut {NoStop}%
\bibitem [{\citenamefont {{Shan}}\ \emph {et~al.}(2021)\citenamefont {{Shan}},
  \citenamefont {{Lackner}}, \citenamefont {{Han}}, \citenamefont {{Sedov}},
  \citenamefont {{Rupprecht}}, \citenamefont {{Knopf}}, \citenamefont
  {{Eilenberger}}, \citenamefont {{Yumigeta}}, \citenamefont {{Watanabe}},
  \citenamefont {{Taniguchi}}, \citenamefont {{Klembt}}, \citenamefont
  {{H{\"o}fling}}, \citenamefont {{Kavokin}}, \citenamefont {{Tongay}},
  \citenamefont {{Schneider}},\ and\ \citenamefont
  {{Ant{\'o}n-Solanas}}}]{Shan2021}%
  \BibitemOpen
  \bibfield  {author} {\bibinfo {author} {\bibfnamefont {H.}~\bibnamefont
  {{Shan}}}, \bibinfo {author} {\bibfnamefont {L.}~\bibnamefont {{Lackner}}},
  \bibinfo {author} {\bibfnamefont {B.}~\bibnamefont {{Han}}}, \bibinfo
  {author} {\bibfnamefont {E.}~\bibnamefont {{Sedov}}}, \bibinfo {author}
  {\bibfnamefont {C.}~\bibnamefont {{Rupprecht}}}, \bibinfo {author}
  {\bibfnamefont {H.}~\bibnamefont {{Knopf}}}, \bibinfo {author} {\bibfnamefont
  {F.}~\bibnamefont {{Eilenberger}}}, \bibinfo {author} {\bibfnamefont
  {K.}~\bibnamefont {{Yumigeta}}}, \bibinfo {author} {\bibfnamefont
  {K.}~\bibnamefont {{Watanabe}}}, \bibinfo {author} {\bibfnamefont
  {T.}~\bibnamefont {{Taniguchi}}}, \bibinfo {author} {\bibfnamefont
  {S.}~\bibnamefont {{Klembt}}}, \bibinfo {author} {\bibfnamefont
  {S.}~\bibnamefont {{H{\"o}fling}}}, \bibinfo {author} {\bibfnamefont {A.~V.}\
  \bibnamefont {{Kavokin}}}, \bibinfo {author} {\bibfnamefont {S.}~\bibnamefont
  {{Tongay}}}, \bibinfo {author} {\bibfnamefont {C.}~\bibnamefont
  {{Schneider}}}, \ and\ \bibinfo {author} {\bibfnamefont {C.}~\bibnamefont
  {{Ant{\'o}n-Solanas}}},\ }\href@noop {} {\bibfield  {journal} {\bibinfo
  {journal} {arXiv e-prints}\ ,\ \bibinfo {eid} {arXiv:2103.10459}} (\bibinfo
  {year} {2021})},\ \Eprint {http://arxiv.org/abs/2103.10459} {arXiv:2103.10459
  [cond-mat.mes-hall]} \BibitemShut {NoStop}%
\bibitem [{\citenamefont {{Wasak}}\ \emph {et~al.}(2021)\citenamefont
  {{Wasak}}, \citenamefont {{Pientka}},\ and\ \citenamefont
  {{Piazza}}}]{Wasak2021}%
  \BibitemOpen
  \bibfield  {author} {\bibinfo {author} {\bibfnamefont {T.}~\bibnamefont
  {{Wasak}}}, \bibinfo {author} {\bibfnamefont {F.}~\bibnamefont {{Pientka}}},
  \ and\ \bibinfo {author} {\bibfnamefont {F.}~\bibnamefont {{Piazza}}},\
  }\href@noop {} {\bibfield  {journal} {\bibinfo  {journal} {arXiv e-prints}\
  ,\ \bibinfo {eid} {arXiv:2103.14040}} (\bibinfo {year} {2021})},\ \Eprint
  {http://arxiv.org/abs/2103.14040} {arXiv:2103.14040 [cond-mat.mes-hall]}
  \BibitemShut {NoStop}%
\bibitem [{\citenamefont {Paik}\ \emph {et~al.}(2019)\citenamefont {Paik},
  \citenamefont {Zhang}, \citenamefont {Burg}, \citenamefont {Gogna},
  \citenamefont {Tutuc},\ and\ \citenamefont {Deng}}]{Paik2019}%
  \BibitemOpen
  \bibfield  {author} {\bibinfo {author} {\bibfnamefont {E.~Y.}\ \bibnamefont
  {Paik}}, \bibinfo {author} {\bibfnamefont {L.}~\bibnamefont {Zhang}},
  \bibinfo {author} {\bibfnamefont {G.~W.}\ \bibnamefont {Burg}}, \bibinfo
  {author} {\bibfnamefont {R.}~\bibnamefont {Gogna}}, \bibinfo {author}
  {\bibfnamefont {E.}~\bibnamefont {Tutuc}}, \ and\ \bibinfo {author}
  {\bibfnamefont {H.}~\bibnamefont {Deng}},\ }\href {\doibase
  10.1038/s41586-019-1779-x} {\bibfield  {journal} {\bibinfo  {journal}
  {Nature}\ }\textbf {\bibinfo {volume} {576}},\ \bibinfo {pages} {80}
  (\bibinfo {year} {2019})}\BibitemShut {NoStop}%
\bibitem [{\citenamefont {Baas}\ \emph {et~al.}(2004)\citenamefont {Baas},
  \citenamefont {Karr}, \citenamefont {Eleuch},\ and\ \citenamefont
  {Giacobino}}]{Baas2004}%
  \BibitemOpen
  \bibfield  {author} {\bibinfo {author} {\bibfnamefont {A.}~\bibnamefont
  {Baas}}, \bibinfo {author} {\bibfnamefont {J.~P.}\ \bibnamefont {Karr}},
  \bibinfo {author} {\bibfnamefont {H.}~\bibnamefont {Eleuch}}, \ and\ \bibinfo
  {author} {\bibfnamefont {E.}~\bibnamefont {Giacobino}},\ }\href {\doibase
  10.1103/PhysRevA.69.023809} {\bibfield  {journal} {\bibinfo  {journal} {Phys.
  Rev. A}\ }\textbf {\bibinfo {volume} {69}},\ \bibinfo {pages} {023809}
  (\bibinfo {year} {2004})}\BibitemShut {NoStop}%
\bibitem [{\citenamefont {Para{\"\i}so}\ \emph {et~al.}(2010)\citenamefont
  {Para{\"\i}so}, \citenamefont {Wouters}, \citenamefont {L{\'e}ger},
  \citenamefont {Morier-Genoud},\ and\ \citenamefont
  {Deveaud-Pl{\'e}dran}}]{Paraiso2010}%
  \BibitemOpen
  \bibfield  {author} {\bibinfo {author} {\bibfnamefont {T.~K.}\ \bibnamefont
  {Para{\"\i}so}}, \bibinfo {author} {\bibfnamefont {M.}~\bibnamefont
  {Wouters}}, \bibinfo {author} {\bibfnamefont {Y.}~\bibnamefont {L{\'e}ger}},
  \bibinfo {author} {\bibfnamefont {F.}~\bibnamefont {Morier-Genoud}}, \ and\
  \bibinfo {author} {\bibfnamefont {B.}~\bibnamefont {Deveaud-Pl{\'e}dran}},\
  }\href {\doibase 10.1038/nmat2787} {\bibfield  {journal} {\bibinfo  {journal}
  {Nature Materials}\ }\textbf {\bibinfo {volume} {9}},\ \bibinfo {pages} {655}
  (\bibinfo {year} {2010})}\BibitemShut {NoStop}%
\bibitem [{\citenamefont {Ouellet-Plamondon}\ \emph {et~al.}(2017)\citenamefont
  {Ouellet-Plamondon}, \citenamefont {Sallen}, \citenamefont {Morier-Genoud},
  \citenamefont {Oberli}, \citenamefont {Portella-Oberli},\ and\ \citenamefont
  {Deveaud}}]{Ouellet2017}%
  \BibitemOpen
  \bibfield  {author} {\bibinfo {author} {\bibfnamefont {C.}~\bibnamefont
  {Ouellet-Plamondon}}, \bibinfo {author} {\bibfnamefont {G.}~\bibnamefont
  {Sallen}}, \bibinfo {author} {\bibfnamefont {F.}~\bibnamefont
  {Morier-Genoud}}, \bibinfo {author} {\bibfnamefont {D.~Y.}\ \bibnamefont
  {Oberli}}, \bibinfo {author} {\bibfnamefont {M.~T.}\ \bibnamefont
  {Portella-Oberli}}, \ and\ \bibinfo {author} {\bibfnamefont {B.}~\bibnamefont
  {Deveaud}},\ }\href {\doibase 10.1103/PhysRevB.95.085302} {\bibfield
  {journal} {\bibinfo  {journal} {Phys. Rev. B}\ }\textbf {\bibinfo {volume}
  {95}},\ \bibinfo {pages} {085302} (\bibinfo {year} {2017})}\BibitemShut
  {NoStop}%
\bibitem [{\citenamefont {Ciuti}\ \emph {et~al.}(2000)\citenamefont {Ciuti},
  \citenamefont {Schwendimann}, \citenamefont {Deveaud},\ and\ \citenamefont
  {Quattropani}}]{Cuiti2000}%
  \BibitemOpen
  \bibfield  {author} {\bibinfo {author} {\bibfnamefont {C.}~\bibnamefont
  {Ciuti}}, \bibinfo {author} {\bibfnamefont {P.}~\bibnamefont {Schwendimann}},
  \bibinfo {author} {\bibfnamefont {B.}~\bibnamefont {Deveaud}}, \ and\
  \bibinfo {author} {\bibfnamefont {A.}~\bibnamefont {Quattropani}},\ }\href
  {\doibase 10.1103/PhysRevB.62.R4825} {\bibfield  {journal} {\bibinfo
  {journal} {Phys. Rev. B}\ }\textbf {\bibinfo {volume} {62}},\ \bibinfo
  {pages} {R4825} (\bibinfo {year} {2000})}\BibitemShut {NoStop}%
\end{thebibliography}%
\maketitle

\begin{widetext}

\section{Master equation and self-consistent approach}
We start discussing the case of incoherent injection of photons. For simplicity, we first neglect saturation effects, which are discussed in detail in Sec.~\ref{AppendixC}. The evolution of the cavity mode, determined by the Heisenberg equation of motion is
\begin{gather}
i\frac{d\hat a}{d t}=-\Delta\omega_c\hat a+\frac{\Omega}{\sqrt{N}}\sum_{i}\hat x_i-i\frac{\gamma_c}{2}\hat a+F,
\label{EqPh}
\end{gather}
now, it is convenient to re-write the cavity mode as $\langle \hat a\rangle=\sqrt{N}\alpha.$ In terms of this parameter, the steady-state can be written as
\begin{gather}
\alpha=\frac{1}{\Delta_c}\left(\frac{F}{\sqrt{N}}+\frac{\Omega}{N}\sum_{i}\langle \hat x_i\rangle\right)=\frac{1}{\Delta_c}(f+\Omega\langle x\rangle),
\label{alpha}
\end{gather}
where $\Delta_c=\Delta\omega_c+i\gamma_c/2$  and $f=F/\sqrt{N}$.

We use the solution for the steady-state of the photon field to define an effective Hamiltonian for the excitons, 
\begin{gather}
\hat H_{\text{local}}=\hat H_{X}+\sum_i(f_\Omega\hat x_i^\dagger+f_\Omega^*\hat x_i),
\end{gather}
where 
\begin{gather}
f_\Omega=\frac{\Omega}{\Delta_c}(f+\Omega\langle x\rangle),
\label{fOmega}
\end{gather}
here, $f_\Omega$ can be regarded as an effective exciton driving term. For relevant experimental parameters, the kinetic energy is heavily suppressed, that is, $U_X\gg t$. Therefore, we treat  exciton tunnelling at the mean-field level, $$-\sum_{ij}t_{ij}\hat x_i^\dagger\hat x_j\rightarrow -J\sum_i (\hat x_i^\dagger \langle x\rangle+ \hat x_i \langle x^\dagger\rangle)$$ with $J=zt,$ being $z$ the coordination number and $t$ the hopping coefficient. Then, we re-define $f_X$ to include at the mean-field level the hopping ratio $t$  
 \begin{gather}
 f_X=\frac{\Omega\,f}{\Delta_c}+\left(\frac{\Omega^2}{\Delta_c}-J\right)\langle \hat x\rangle,
 \end{gather}
the effects of exciton tunneling are discussed in detail in section ~\ref{NN}.

The evolution of the density matrix for the excitons is governed by the effective master equation 
\begin{gather}
\frac{d\hat\rho}{dt}=-i[\hat H_{\text{local}},\hat\rho]+\mathcal D_x[\hat\rho]=\mathcal L_x[\hat\rho],
\label{eqSs}
\end{gather}
where we should recall that the Hamiltonian $\hat H_{\text{local}}$ depends itself on the expectation $\langle \hat x\rangle=\text{Tr}(\hat x\hat\rho).$ Therefore, we adapt self-consistently the Lindblad operator. The self-consistent approach, which iterates on $\langle \hat x\rangle$ can be summarised as follows. We numerically diagonalise the local Lindblad operator taking a given value of  $\langle \hat x\rangle^{(i)},$ where $i$ denotes the $i$-th iteration step, which is employed in  Eq.~\ref{fOmega} to define the effective exciton Hamiltonian. We obtain the steady-state $\hat\rho^{(i)}$ which we use to evaluate the expectation value of $\hat x$. This defines $\langle \hat x\rangle^{(i+1)}=\text{Tr}(\hat x\hat\rho^{(i)}).$ We then replace $\langle \hat x\rangle^{(i+1)}$ in Eq.~\ref{fOmega} and repeat the iterative scheme. Numerically, we diagonalise the local Lindblad operator for $N_L=225$ corresponding to a cut-off for the exciton's level of $N=15$ and 200 iteration cycles unless stated explicitly otherwise. 

For $\Omega_{\text{sat}}=0,$ our numerics can be benchmarked against the analytical solutions which can be extended from the case of an isolated cavity with a Kerr-like non-linearity~\cite{Drummond1980}. The density matrix extended from ~\cite{Drummond1980}, can be written as a self-consistent equation
\begin{gather}
\label{DMatrix}
\hat \rho_{n,m}=\frac{(-2)^{n+m}}{\sqrt{n!m!}}\left(\frac{f_X}{U_X}\right)^n\left(\frac{f_X^*}{U_X}\right)^m\frac{\Gamma\left(\frac{-2\Delta_X}{U_X}\right)\Gamma\left(\frac{-2\Delta_X^*}{U_X}\right)}{\Gamma\left(\frac{-2\Delta_X}{U_X}+n\right)\Gamma\left(\frac{-2\Delta_X^*}{U_X}+m\right)}\frac{_0\mathcal F^{2}\left(\frac{-2\Delta_X}{U_X}+n,\frac{-2\Delta_X^*}{U_X}+m,4\left|\frac{f_x}{U_X}\right|^2\right)}{_0\mathcal F^{2}\left(\frac{-2\Delta_X}{U_X},\frac{-2\Delta_X^*}{U_X},8\left|\frac{f_X}{U_X}\right|^2\right)},
\end{gather}
where $_0\mathcal F^{2}(a,b;z)$ is the hypergeometric function defined as
\begin{gather}
_0\mathcal F^{2}(a,b;z)=\sum_{n=0}^{\infty}\frac{\Gamma(a)\Gamma(b)}{\Gamma(a+n)\Gamma(b+n)}\frac{z^n}{n!},
\end{gather}
here, the self-consistency stems from the dependance of $f_X$ on $\text{Tr}(\hat\rho\,\hat x).$  Before comparing these approaches, we shortly provide details on the extended Gross-Pitaeksvii equation.
\subsection{Extended Gross-Pitaevskii equation and master equation}
We now turn our attention into the comparison between the semi-classical description of the excitons, the numerical self-consistent scheme, the density matrix in Eq.~\ref{DMatrix} obtained self-consistenly, and the non-interacting case. First, we consider non-interacting excitons $U_X=0$, in this case, the steady-state acquires the simple form of
\begin{gather}
\alpha=\frac{f}{\Delta_c}\frac{1}{1-\frac{\Omega^2}{\Delta_x\Delta_c}},\\ 
\langle \hat x \rangle=\frac{\Omega}{\Delta_x}\alpha.
\label{idealSS}
\end{gather}
On the other hand, after some algebra and introducing $\phi=\langle \hat x \rangle,$ the  extended driven-dissipative Gross-Pitaevskii for the steady-state reads can be written as
\begin{gather}
\left(-\Delta_X-J+\frac{\Omega^2}{\Delta_c}+U_X|\phi|^2\right)\phi+\frac{\Omega\,f}{\Delta_c}=0.
\end{gather}
We can now compare the coherence amplitude $|\phi|$ predicted from the different approximations. In Fig.~\ref{FigA1} we plot the coherence amplitude, the red squares correspond to the self-consistent approach described for the Lindblad operator $\mathcal L,$  while the solid blue line gives the benchmark for the density matrix in  Eq.~\ref{DMatrix}. This excellent agreement remarks the equivalence between these two numerical approaches. The black circles correspond to the eGP equation which predicts a bi-stability. Finally, the solid green line gives the solution for non-interacting excitons, namely Eq.~\ref{idealSS}. As discussed in the main text, for low intensities, the coherence amplitude is governed by Eq.~\ref{idealSS}. As soon as the interactions become relevant, the coherence amplitude obtained from Eq.~\ref{DMatrix} completely deviates from the eGP prediction illustrating the strongly interacting nature of the excitons, which cannot be explained by the eGP equation.

\begin{figure}[ht]
\begin{center}
\includegraphics[width=0.6\columnwidth]{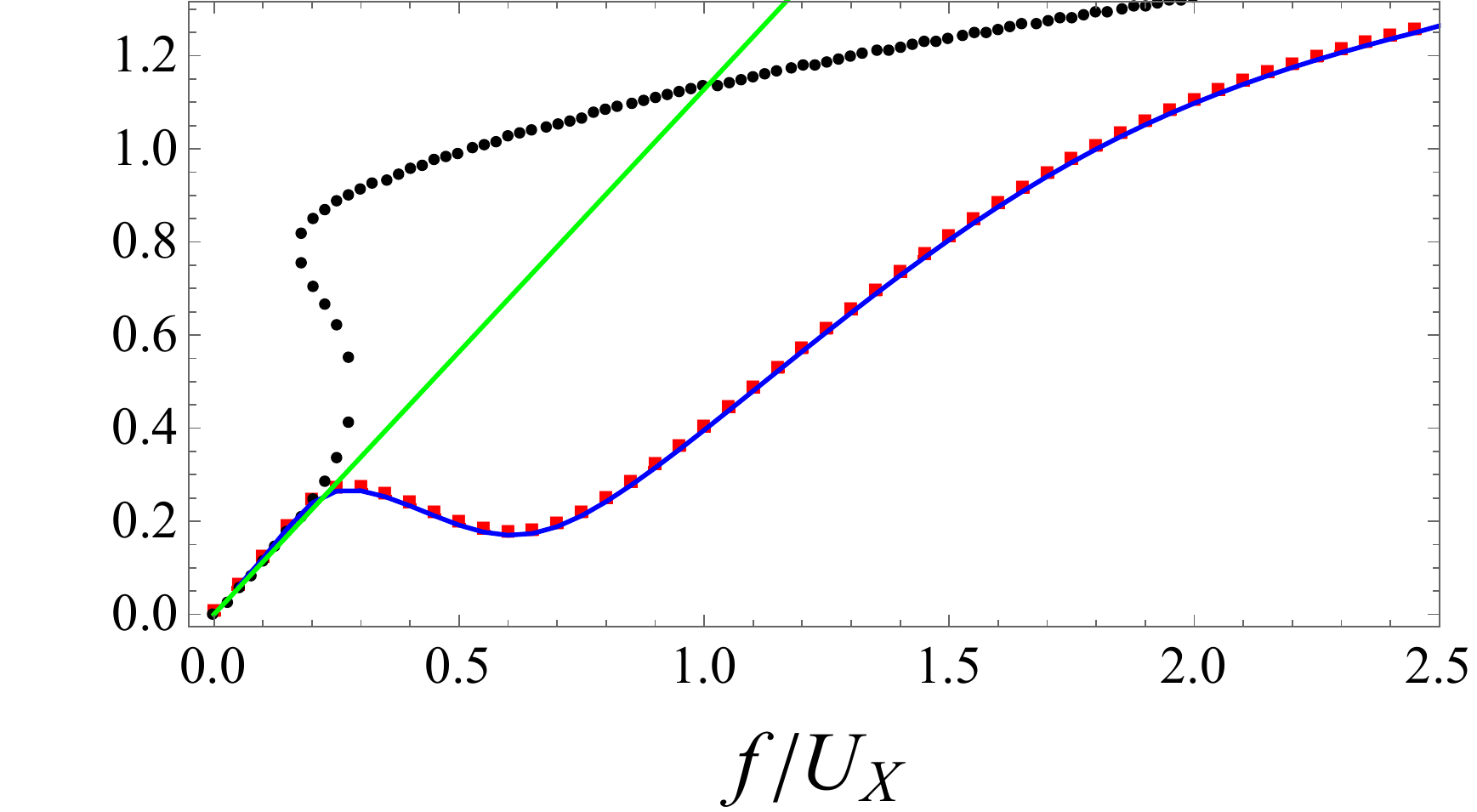}
\end{center}
\caption{Comparison between different approaches for the same value of the parameters than Fig.~3 (main text). The solid green line depicts the steady state for non-interacting excitons, the black circles correspond to the eGP equation, the red squares and solid blue line illustrate the equivalence between the self-consistent diagonalisation of the Lindblad operator and the iterative approach for the reduced density matrix in Eq.~\ref{DMatrix} respectively.} 
\label{FigA1}
\end{figure} 

\section{Nearest neighbour tunnelling and Saturation effects}

\subsection{Nearest neighbour tunnelling effects} 
\label{NN}
The hybridization of the excitonic band is accompanied by a strong suppression of their kinetic energy, this leads to a small hopping coefficient $t\ll U_X$ which can be understood at the mean-field level. 

The corrections due to nearest-neighbour tunnelling (NN) are expected to be small for
\begin{gather}\left|\frac{\Omega^2}{\Delta_c}\right|\gg zt=J,
\end{gather}
that is, as long as the second-order mediated tunnelling coefficient remains larger than the NN hopping ratio, we expect the tunnelling effects to be governed by the light-mediated tunnelling.  

To provide quantitative estimates of the NN hopping we consider a hopping coefficient $J/U_X=0.3.$  In this case, the ratio between the tunnelling amplitudes is  $\left|\frac{\Omega^2}{\Delta_c}\right|/J\approx 4$, for the light-matter coupling exemplified in the main text. In Fig.~\ref{FigAB1} we show the effects of NN tunnelling for a negative photon detuning $\Delta\omega_c=-1$ (red squares) compared against vanishing tunneling $J/U_X=0$ (blue circles). Fig.~\ref{FigAB1} shows that in this regime, the main features are only slightly affected by the NN hopping, in particular, the tunnelling mediated bi-stability is slightly enhanced by additional terms contributing to the hopping.

\begin{figure}[h!]
\begin{center}
\includegraphics[width=0.6\columnwidth]{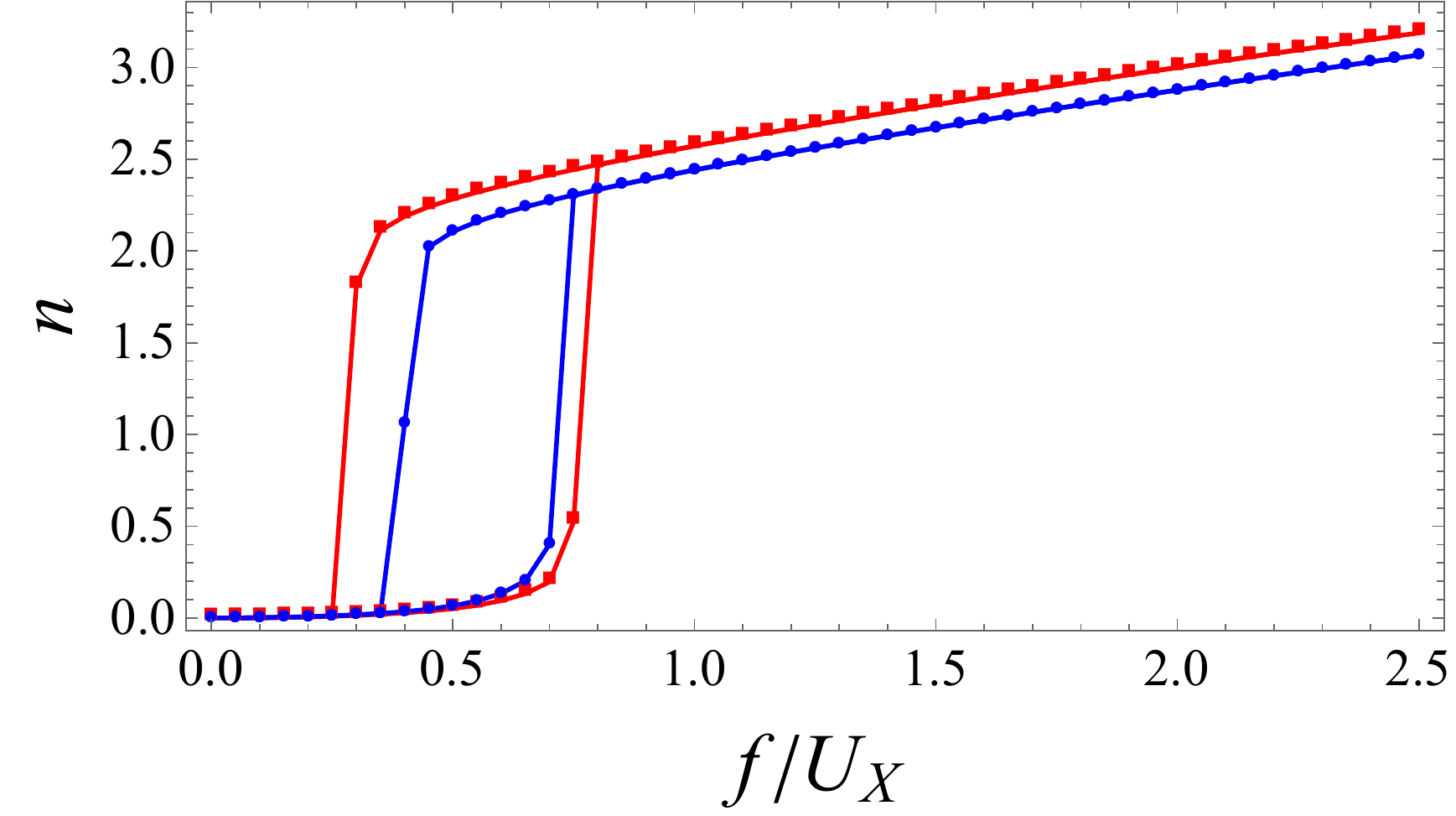}
\end{center}
\caption{Hysteresis and lower branch for $J/U_X=0.3$ (red squares) and $J/U_X=0$ (blue circles), which explicitly shows that the hysteresis and sudden jump are robust in presence of direct NN exciton tunnelling.  } 
\label{FigAB1}
\end{figure} 

\label{AppendixB}
\subsection{Saturation effects}
\label{AppendixC}
Saturation effects are included via an anharmonic light-matter coupling, such that the light-matter coupling reads as
\begin{gather}
\hat H_{l-m}=\frac{1}{\sqrt{N}}\sum_i \hat a^\dagger (\Omega-\Omega_{\text{sat}}\hat n_i)\hat x_i+\hat x_i^\dagger (\Omega-\Omega_{\text{sat}}\hat n_i)\hat a=\\ \nonumber
=\frac{\Omega}{\sqrt{N}}\sum_i \hat a^\dagger \left(1-\frac{\hat n_i}{n_\text{sat}}\right)\hat x_i+\hat x_i^\dagger \left(1-\frac{\hat n_i}{n_\text{sat}}\right)\hat a,
\end{gather}
the anharmonic term can be understood in terms of the suppression of the light-matter coupling $\Omega\rightarrow \Omega-\Omega_{\text{sat}}\hat n$. Then, saturation effects are relevant for densities of the order   $\langle \hat n \rangle \approx n_\text{sat}=\Omega_{\text{sat}}/\Omega.$ On the other hand, for  $ \langle n \rangle \ll n_\text{sat}$ the reduction of the light-matter coupling can safely be ignored.

For two-dimensional gases, the reduction of the light-matter coupling $\Omega_{\text{sat}}=\Omega/(n_\text{sat}A)$ given in terms of the saturation density $n_{\text{sat}}=7/(16\pi a_B^2)$  with $a_B$ the exciton radius and the macroscopic area $A$~\cite{Cuiti2000} is a reminder that the intrinsic nature of the excitons becomes relevant when the inter-particle distance between excitons is of the order of the Bohr exciton radius. The spatial localisation of the hybrid excitons to the moir\'e lattice sites modifies the saturation condition but follows the same physical argument. However, the macroscopic area $A$ should be replaced by the ``local area" of the lattice site $A_M$. The local area is obtained by assuming that the spatial localisation of the excitons can be described by a periodic potential that can be approximated as a parabolic trap around each lattice site. Thus, around the minima of the lattice site, we consider a potential of the form $\frac{m \omega_M r_i^2}{2}.$ Therefore, the site dynamics of the exciton is governed by an harmonic oscillator Hamiltonian $$\hat H_{ho}=\frac{\hat p^2}{2m}+\frac{m \omega_M^2 \hat r^2}{2}.$$ For the ground-state, the classical return point $r_s=\frac{1}{\sqrt{m\omega_M}}$ defines the characteristic extension of the exciton wave function. Here, $m$ is the exciton mass, and $\omega_M $ is proportional to the electron tunnelling amplitude, responsible for the moir\'e lattice, typically of the order of $\omega_M=5-20\text{meV}$~\cite{Zhang2021}. We estimate  $r_s\approx 1-5\text{nm},$ whereas the typical exciton radius is $a_B\approx1\text{nm}$. The saturation density remains inversely proportional to the square of the exciton Bohr radius $n_s\propto 1/a_B^2$. Thus, the reduced light-matter coupling is estimated by $\Omega_{\text{sat}}/\Omega\propto (a_B/r_s)^2 \in \{0.1,1\}$. The suppression of the light-matter coupling sensible to one exciton per site is consistent with the experimental observation of moir\'e polaritons~\cite{Zhang2021}. This illustrates  the stark contrast between two-dimensional polariton gases and moir\'e polaritons which exhibit additional non-linearities even at the level of one and a few excitons per site~\cite{Zhang2021}. Here we take units where $\hbar=1$ and take $m=0.8m_e$ being $m_e$ the bare electron mass.

The anharmonic light-matter coupling modifies both the steady-state of the cavity mode, and the effective Hamiltonian for the excitons. The steady-state of the cavity acquires an explicit dependence on $\langle \hat n\hat x\rangle$  and reads as
\begin{gather}
\alpha=\frac{1}{\Delta_c}(f+\Omega\langle \hat x\rangle-\Omega_{\text{sat}}\langle \hat n\hat x\rangle),
\end{gather}
thus in addition to the dependence on the coherence $\phi=\langle \hat x\rangle$, the Hamiltonian develops an extra term consequence of the saturation of the exciton-photon coupling. In general, we should note that 
\begin{gather}
\Omega\langle \hat x\rangle-\Omega_{\text{sat}}\langle \hat n\hat x\rangle\neq (\Omega-\langle n\rangle\Omega_{\text{sat}})\langle\hat x\rangle,
\label{wrong}
\end{gather} thus we explicitly evaluate $\langle \hat n\hat x\rangle$. The right-side of Eq.~\ref{wrong} gives, however, intuition to understand the physical meaning of the saturation term. 
The inclusion of the anharmonic light-matter coupling renders the approach based on the analytical expression for the density matrix in Eq.~\ref{DMatrix} no longer valid. Therefore, the density matrix is now only obtained by means of the self-consistent method for the Lindblad super-operator.

To complement our discuss from the main text we show in Fig.~\ref{FigAC1} the coherence amplitude and purity as a function of the pump intensity and exciton detuning for the same parameters than in Fig.~2 (e) (main text). We show that for the low-densities, the lobular pattern persists and follows the discussion in the main text, where the transition from negligible site occupation to a finite number of exciton per site is accompanied by a drop of the purity and a non-monotonous behaviour of the coherence through the crossover.  In Fig.~\ref{FigAC1} we employ 100 iteration cycles and for the hysteresis branch we drive the system from an initial $f/\Omega=6$ (from above).

\begin{figure}[h!]
\begin{center}
\includegraphics[width=.65\columnwidth]{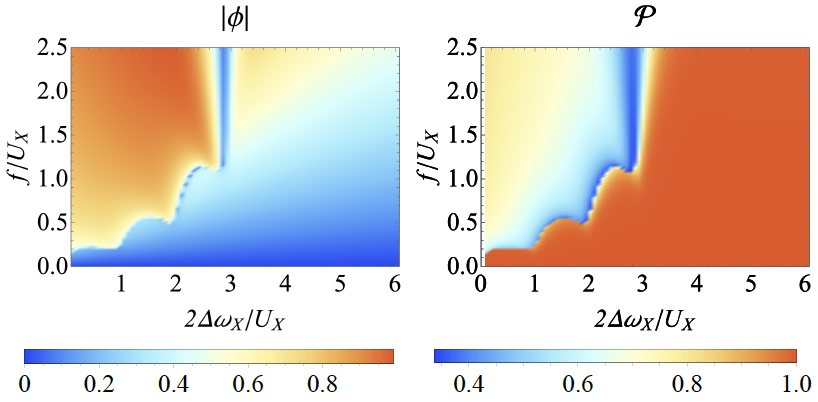}
\end{center}
\caption{Steady-state properties for the lower-branch for large saturation effects $\Omega_{\text{sat}}=\Omega/4.$ (Left) Coherence amplitude. (Right) Purity. We employ the same parameters than in Fig.~2 (e) (main text). } 
\label{FigAC1}
\end{figure} 

For negative exciton detuning, a bi-stable phase persists even when saturation effects become relevant,  the bi-stability is illustrated by cross-sections of the exciton number for fixed exciton detuning and as a function of $f/U_X$ in Fig.~\ref{FigAC2} for  $\Omega_{\text{sat}}=\Omega/4$, the red squares correspond to the hysteresis branch, and the blue circles to a $f/\Omega$ tuned from below.  Here we illustrate the persistence of the bi-stability for $\Delta\omega_X/U_X=0.6$ (left) and  $\Delta\omega_X/U_X=1.25$ (right).
\begin{figure}[h!]
\begin{center}
\includegraphics[width=0.35\columnwidth]{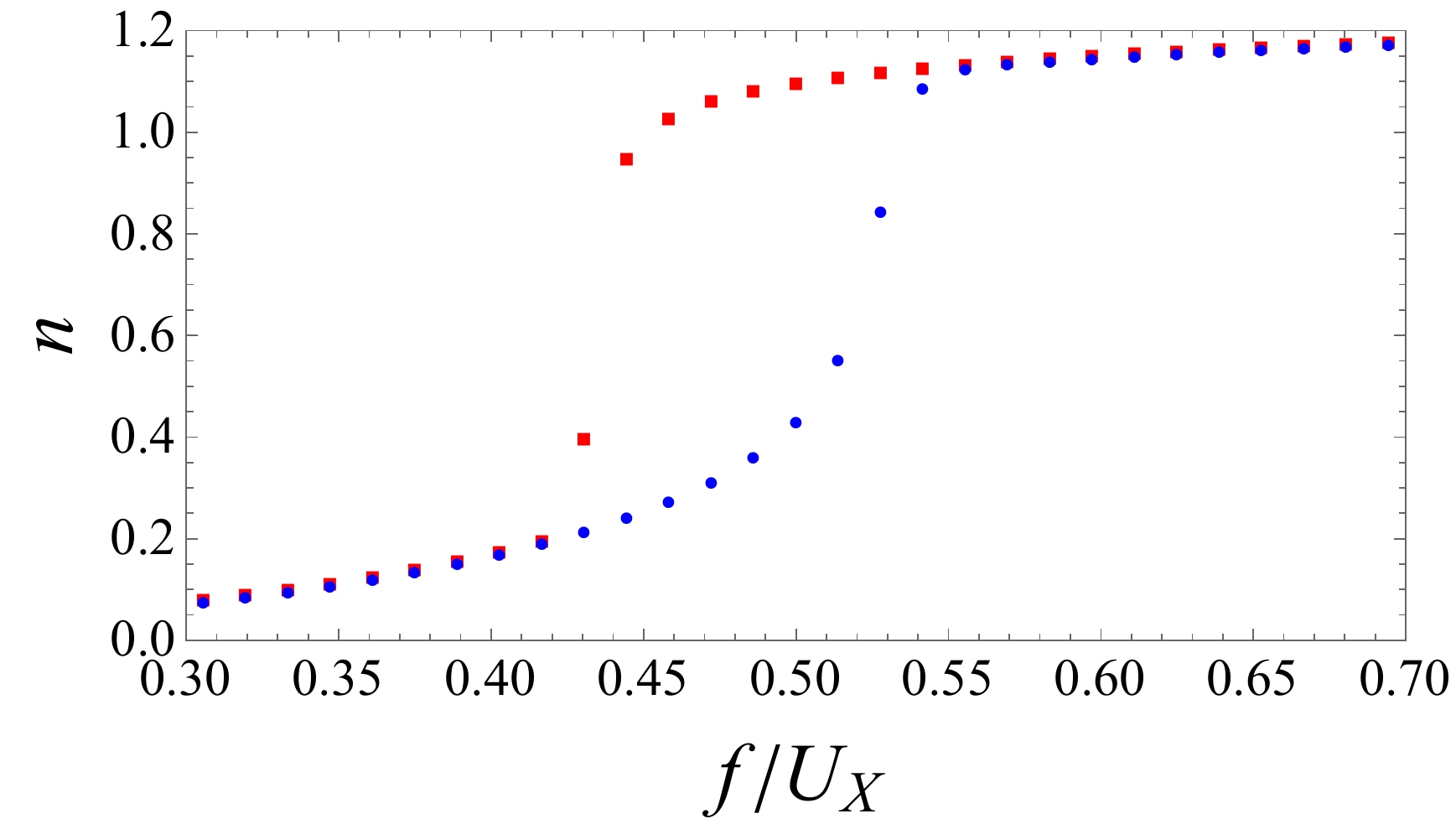}
\includegraphics[width=0.35\columnwidth]{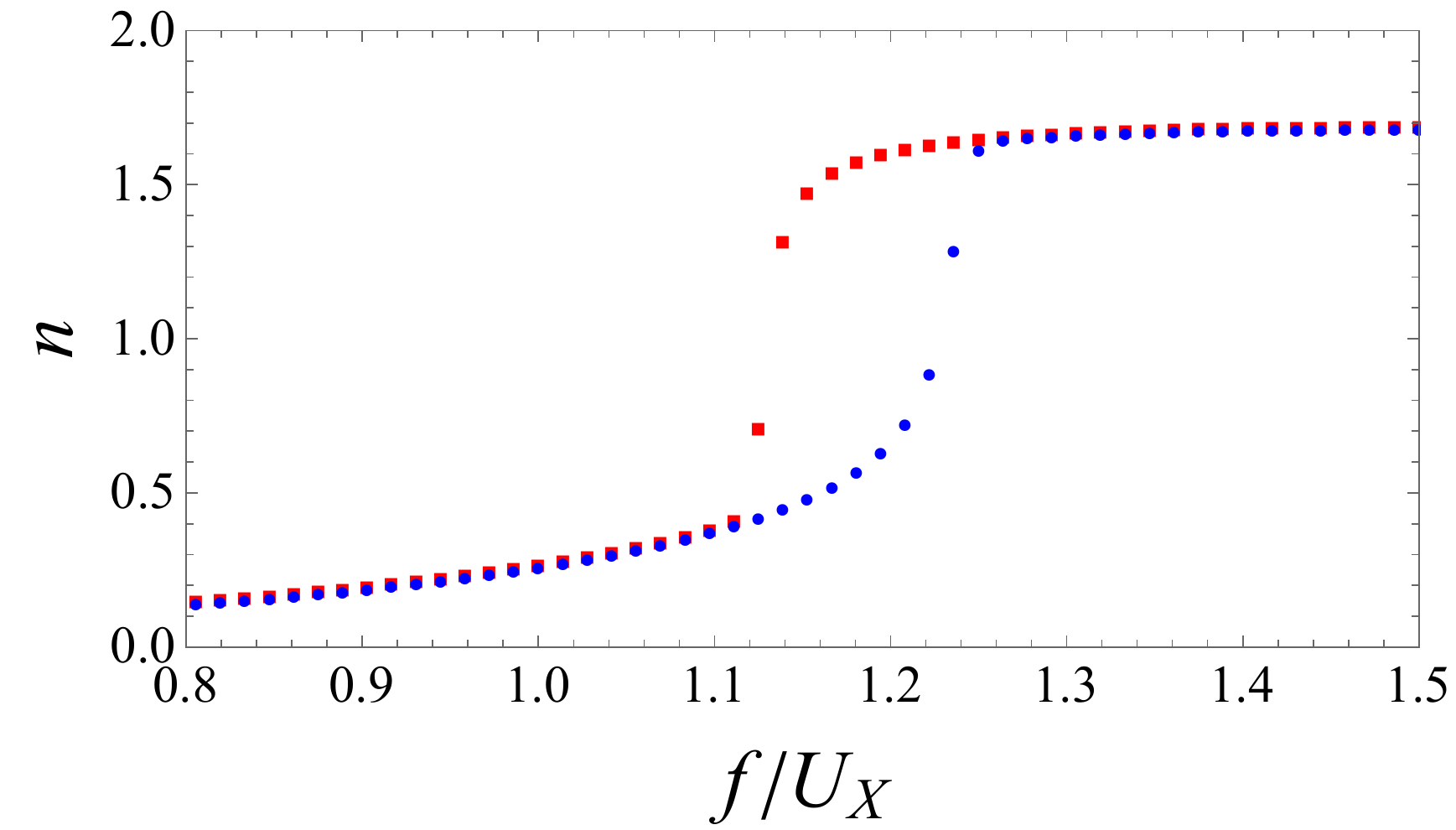}
\end{center}
\caption{Bi-stability  for $\Delta\omega_X/U_X=0.6$ (left) and  $\Delta\omega_X/U_X=1.25$ (right) and saturation effects of the order of $\Omega_{\text{sat}}=\Omega/4$. (Red squares depict the hysteresis branch, while blue circles correspond to the lower branch. } 
\label{FigAC2}
\end{figure} 

Finally, in Fig.~\ref{FigAC3} we compare the exciton number for different values of $\Omega_{\text{sat}}$ for $\Delta\omega_X/U_X=1.25,$ the hysteresis branch.  The red squares correspond to $\Omega_{\text{sat}}/\Omega=0.1$, the blue circles $\Omega_{\text{sat}}/\Omega=0.25$, and the black triangles $\Omega_{\text{sat}}/\Omega=0.5$. In this case, even for $\Omega_{\text{sat}}/\Omega=0.25$ saturation effects become prominent, the abrupt jump for small   $\Omega_{\text{sat}}/\Omega$ is visibly smoothen, the exciton number decreases, and the jump is shifted to larger values of $f/\Omega$. 
\begin{figure}[h!]
\begin{center}
\includegraphics[width=0.35\columnwidth]{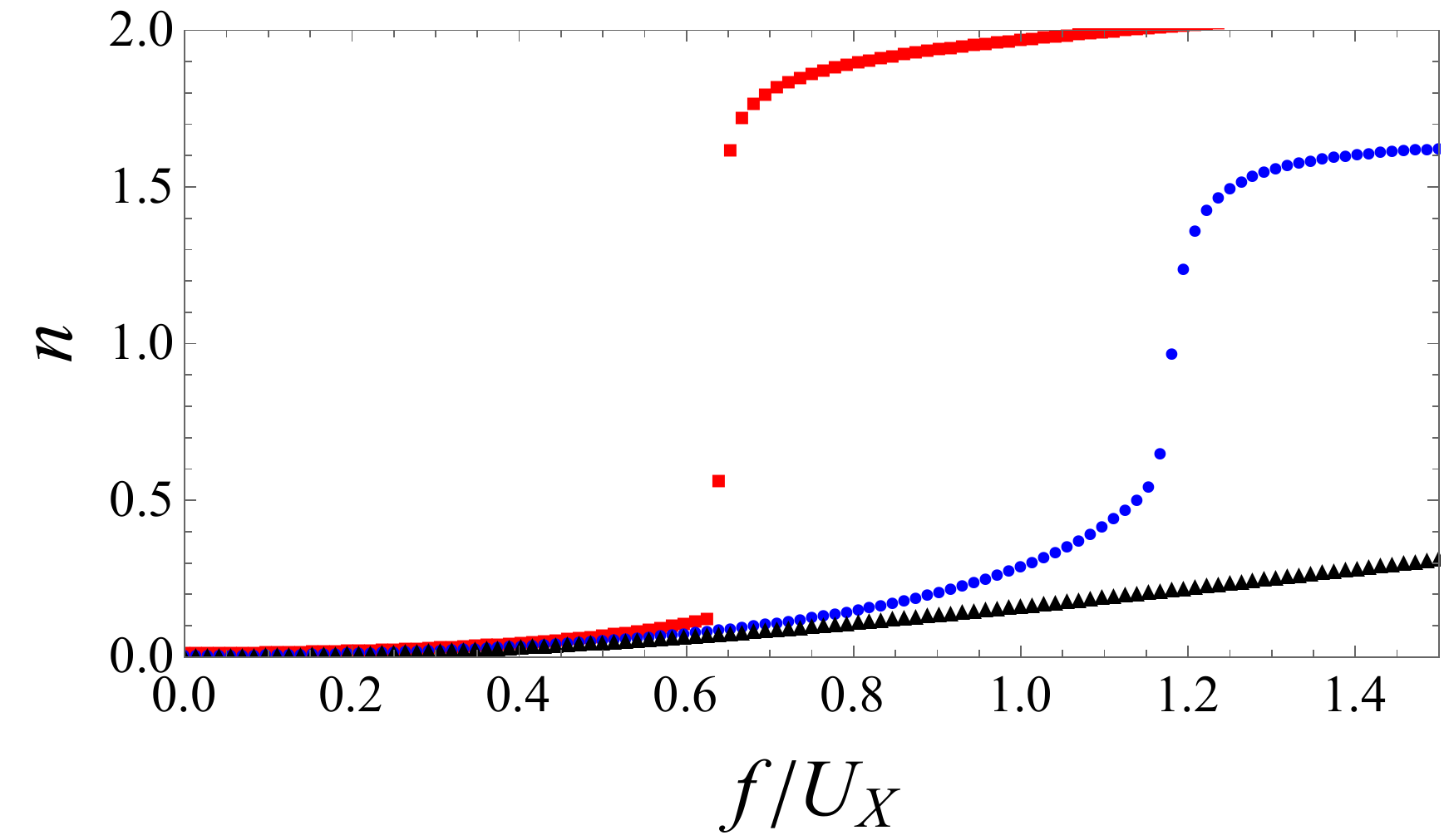}
\end{center}
\caption{Exciton number for several values of $\Omega_{\text{sat}}$.  The red squares correspond to $\Omega_{\text{sat}}/\Omega=0.1$, the blue circles to $\Omega_{\text{sat}}/\Omega=0.25$, and the black triangles to $\Omega_{\text{sat}}/\Omega=0.5$ } 
\label{FigAC3}
\end{figure} 
\section{Lasers: From single-photon to multi-photon resonances}
\subsection{Effective two-level system}
Let us turn our attention into the details on the single- and multi-photon resonances in the scenario with exciton gain. For simplicity, we start discussing the limit where only empty moir\'e lattice sites can be driven, in this case, the gain and losses processes are described by\begin{gather}
\mathcal D_x[\hat\rho]=\Gamma_x\mathcal D[\hat y^\dagger]\hat\rho+\gamma_x\mathcal D[\hat x]\hat \rho,
\end{gather}
here, the first term accounts for the driving of empty sites, such that $\hat y$ and $\hat y^\dagger$ are exciton operators restricted to $N_x=0,1$ subspace. Now, we solve the equation $\mathcal L[\hat\rho]=0$ in a rotating frame with frequency $\omega.$ We define $\Delta\omega_X=\omega-\omega_X$ and $\Delta\omega_c=\omega-\omega_c$, where the photon energy $\omega$ is also left to be obtained. In the limit of $U_X\gg \Omega,$ the system can be regarded as a two-level system with  equations for the steady-state
\begin{gather}
0=\gamma \rho_{11}+\frac{4 \gamma_c\rho_{01}\rho_{10} \Omega ^2}{\gamma_c^2+4 (\delta +\omega )^2}-\Gamma_x\rho_{00},\\
0=\frac{2\rho_{01} \Omega ^2 (\rho_{11}-\rho_{00})}{\gamma_c+2 i (\delta +\omega )}-\frac{1}{2}\rho_{01} (\gamma +\Gamma_x+2 i  \omega ),\\ 
0=\frac{2\rho_{10} \Omega ^2 (\rho_{11}-\rho_{00})}{\gamma_c-2 i (\delta + \omega )}-\frac{1}{2}\rho_{10} (\gamma +\Gamma_x-2 i  \omega ),
\end{gather}
with $\delta=\omega_X-\omega_c$. In this case, there is a trivial solution with $\rho_{00}=\gamma_x/(\gamma_x+\Gamma_x)$ and $\rho_{01}=\rho_{10}=0$. A non-trivial solution appears for $\Omega>\Omega_c$ and $\Gamma_x>\gamma_x$, being $\Omega_c=(\gamma_x+\Gamma_x)\sqrt{\frac{\gamma_c}{4(\Gamma_x-\gamma_x)}},$ with  
\begin{gather}
\rho_{11} =\frac{1}{2}\left(1+\frac{\gamma_c(\gamma_x+\Gamma_x)}{4\Omega^2}\right)+\frac{1}{2}\left(\delta^2\frac{\gamma_c(\gamma_x+\Gamma_x)}{2(\gamma_x+\Gamma_x+\gamma_c)\Omega^2}\right)\\  \label{EnT}
\rho_{01}=\frac{i \sqrt{\gamma_c} \sqrt{(\gamma +\gamma_c+\Gamma_x)^2+4 \delta ^2} \sqrt{\gamma_c (\gamma +\Gamma_x)^2 \left((\gamma +\gamma_c+\Gamma_x)^2+4 \delta ^2\right)+4 \Omega ^2 (\gamma -\Gamma_x) (\gamma +\gamma_c+\Gamma_x)^2}}{4 \sqrt{2} \sqrt{\Omega ^4 (\gamma +\gamma_c+\Gamma_x)^4}},\\ \label{phEr}
\omega=\omega_X-\frac{\gamma_x+\Gamma_x}{\gamma_x+\Gamma_x+\gamma_c}\delta.
\end{gather}

In Fig.~\ref{FigAC4} we show the expectation value of $\hat x$ as a function of $\Omega$ for the two-level system in presence of exciton driving. The red dots correspond to the numerical implementation of the self-consistent approach. The solid blue line illustrates the analytical solution in Eq.~\ref{EnT}, while the dashed black line gives the onset for the non-trivial state. The onset $\Omega_c$ increases with $\gamma_x\rightarrow\Gamma_x$.  In this case, we find the photon energy self-consistently by solving the equivalent of Eq.~\ref{EqPh}. In general the photon energy slightly deviates from $\omega_c$ as shown by Eq.~\ref{phEr} depending on the cavity detuning $\delta$, and the damping and driving rates.

\begin{figure}[h!]
\begin{center}
\includegraphics[width=0.6\columnwidth]{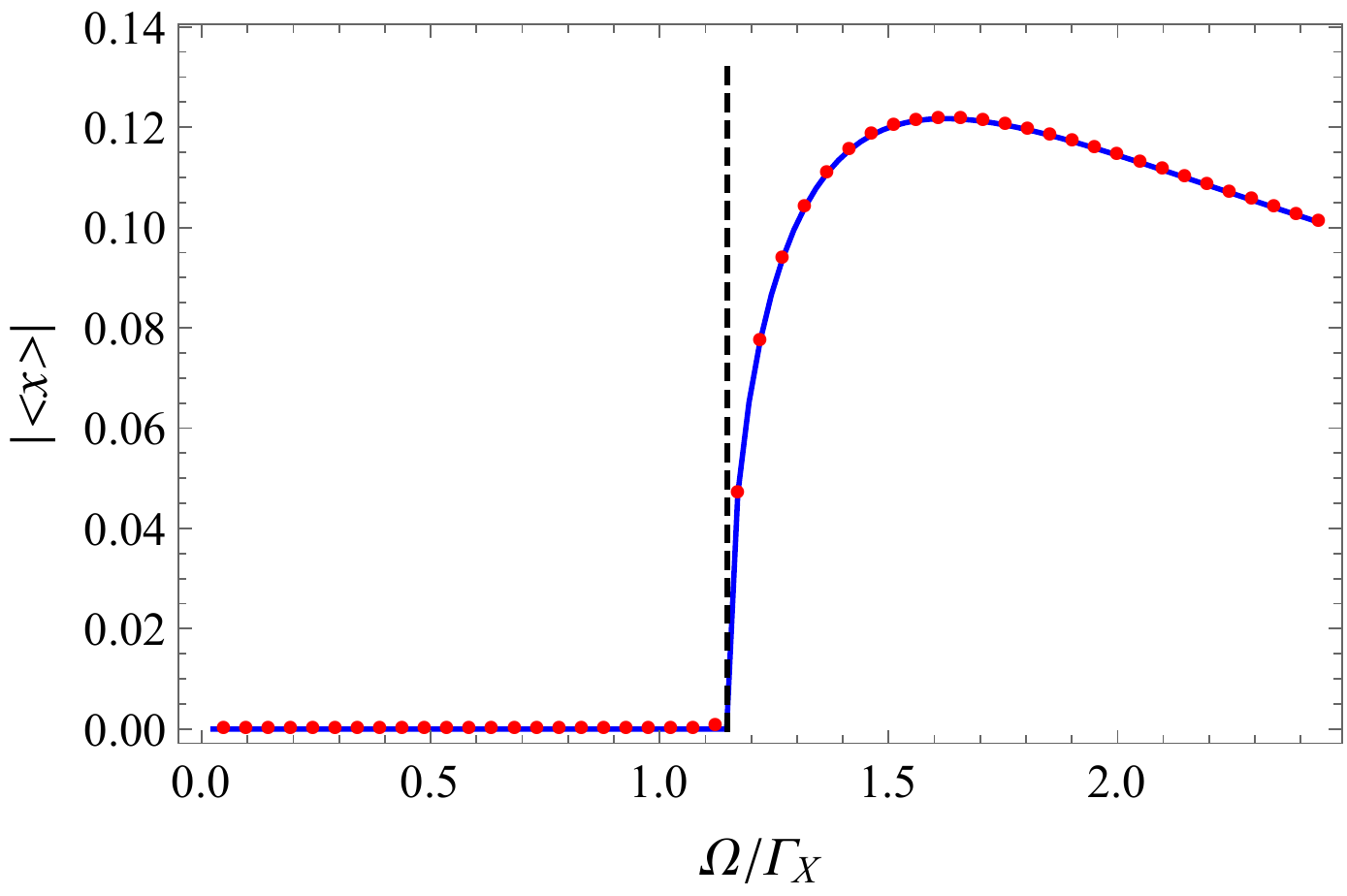}
\end{center}
\caption{Expectation value $\langle \hat x\rangle$ as a function of $\Omega$ for $\Gamma_x/\gamma_x=6,$ and $\Gamma_x=\gamma_c$. The red dots correspond to the self-consistent exact diagonalization scheme, the blue line the analytical solution in Eq.~\ref{EnT}. The vertical black dashed line gives $\Omega_c$. } 
\label{FigAC4}
\end{figure} 

For the two-photon resonance, we cannot longer derive an analytical expression for the photon amplitude, instead, we solve for the Lindblad operator for the exciton field $\mathcal L=\mathcal L(\omega,\alpha),$ where the photon energy $\omega$ and amplitude $\alpha$ are variables obtained self-consistently.  We stress that this self-consistent approach recovers the exact solution for the two-level system in Fig.~\ref{FigAC4} as well as, for the coherent injection of photons in absence of saturation effects, as discussed previously.

\subsection{Three-level effective Hamiltonian: Rate equations}
To understand better the two-photon resonance, we derive the rate equations for an effective three-level system. We start by solving for the off-diagonal elements of the exciton's density matrix, which evolution is given by
\begin{gather}
\frac{d \rho_{01}}{dt}=-i \Omega  \left(\sqrt{2} \alpha \rho_{02}+\alpha^{*} (\rho_{00}-\rho_{11})\right)+\sqrt{2} \gamma_x \rho_{12}+\frac{1}{2} i\rho_{01} (U_X+i (\gamma_x +3 \Gamma_x),\\
\frac{d \rho_{02}}{dt}=i \alpha^{*} \Omega  \left(\rho _{12}-\sqrt{2}\rho_{01}\right)-\gamma_x \rho_{02}-\frac{\Gamma_x\rho_{02}}{2},\\ 
\frac{d \rho_{12}}{dt}=i \Omega  \left(\alpha \rho_{02}+\sqrt{2} \alpha^{*} (\rho_{22}-\rho_{11})\right)-\frac{3 \gamma_x \rho_{12}}{2}+\Gamma_x \left(\sqrt{2}\rho_{01}-\rho_{12}\right)-\frac{i\rho_{12} U}{2}.
\end{gather}
for the steady-state, and after some algebra, we find  a solution for the off-diagonal elements of the density matrix in terms of only its diagonal elements. These solutions are
\begin{gather}
\rho_{01}=\frac{2 \alpha^{*} \Omega  \left(-4 n_\alpha \Omega ^2 (\rho_{00}+\rho_{11}-2\rho_{22})+(2 \gamma_x -\Gamma_x (\gamma  (3\rho_{00}+\rho_{11}-4\rho_{22})+2 \Gamma_x (\rho_{00}-\rho_{11})+i U_X (\rho_{00}-\rho_{11}))\right)}{i (2 \gamma_x -\Gamma_x \left(3 \gamma_x ^2+3 \gamma_x  \Gamma_x+6 \Gamma_x^2+U_X^2-2 i \gamma  U+i \Gamma_x U_X\right)+4 \alpha  \alpha^{*} \Omega ^2 (U_X-3 i (\gamma_x +\Gamma_x)},\\ 
\rho_{12}=\frac{2 i \sqrt{2} \alpha^{*} \Omega  \left(-4 n_\alpha \Omega ^2 (\rho_{00}+\rho_{11}-2\rho_{22})+(2 \gamma_x -\Gamma_x (\Gamma_x (2\rho_{00}+\rho_{11}-3\rho_{22})+(\rho_{11}-\rho_{22}) (\gamma_x -i U_X))\right)}{4 \alpha  \alpha^{*} \Omega ^2 (3 (\gamma_x +\Gamma_X+i U_X)-(2 \gamma_x-\Gamma_x \left(3 \gamma_X ^2+3 \gamma_x  \Gamma_x+6 \Gamma_x^2+U_X^2-2 i \gamma_x  U_X+i \Gamma_x U_X\right)},\\
\rho_{02}=-\frac{4 \sqrt{2} (\alpha^{*})^2 \Omega ^2 (U_X(\rho_{00}-\rho_{22})-3 i (\gamma_x  (\rho_{00}-\rho_{22})+\Gamma_x (\rho_{22}-\rho_{11})))}{i (2 \gamma_x -\Gamma_x \left(3 \gamma_x ^2+3 \gamma_x  \Gamma_x+6 \Gamma_x^2+U^2-2 i \gamma_x U+i \Gamma_x U_X\right)+4 \alpha  \alpha^{*} \Omega ^2 (U_X-3 i (\gamma_x +\Gamma_x)}.
\end{gather}
To make progress, we make some simplifications, first we work in the regime where $\Gamma_x\gg \gamma_x$ and fix $\gamma_x=0,$ then we expand the off-diagonal elements in powers of  $\tilde\Omega=\Omega/U_X$ and retain terms up to $\tilde \Omega^3$. Thus, we obtain
\begin{gather}
\rho_{01}=2 \alpha^{*} \tilde\Omega  \left(\frac{8 n_\alpha (2 \tilde\Gamma_x+i) \tilde\Omega ^2 (-3 i \tilde\Gamma_x (\rho_{11}-\rho_{22})-\rho_{00}+\rho_{22})}{\tilde\Gamma_x (1+\tilde\Gamma_x (6 \tilde\Gamma_x+i))^2}-\frac{i (\rho_{00}-\rho_{11})}{3 \tilde\Gamma_x-i}\right),\\
\rho_{12}=\frac{8 \sqrt{2} n_\alpha\,\alpha^{*} (\tilde\Gamma_x+i) \tilde\Omega ^3 (-3 i \tilde\Gamma_x (\rho_{11}-\rho_{22})-\rho_{00}+\rho_{22})}{\tilde\Gamma_x (1+\tilde\Gamma_x (6 \tilde\Gamma_x+i))^2}+\frac{2 \sqrt{2} \alpha^{*} \tilde\Omega  (-i \tilde\Gamma_x (2\rho_{00}+\rho_{11}-3\rho_{22})-\rho_{11}+\rho_{22})}{1+\tilde\Gamma_x (6 \tilde\Gamma_x+i)},\\
\rho_{02}=\frac{4 \sqrt{2} (\alpha^{*})^2 \tilde\Omega ^2 (3 \tilde\Gamma_x (\rho_{11}-\rho_{22})-i\rho_{00}+i\rho_{22})}{\tilde\Gamma_x (1+\tilde\Gamma_x (6 \tilde\Gamma_x+i))},
\end{gather}
where we introduced $\tilde\Gamma_x=\Gamma_x/U_X.$ For the two-photon process, the single-particle transitions are far detuned, and for two-photon gain purposes, the linear terms in $\tilde \Omega$ can be neglected. One can further simplify these expressions by taking $\tilde \Gamma_x\ll 1$. Then, the elements for the off-diagonal density matrix take the following form
\begin{gather}
\label{g2rho}
\rho_{01}\approx \frac{16 i n_\alpha  \alpha^{*} \tilde\Omega ^3 (\rho_{22}-\rho_{00})}{\tilde\Gamma_x}+16 \alpha^*n_\alpha \tilde\Omega ^3 (-4 \rho_{00}+3 \rho_{11}+\rho_{22})\approx\frac{16 i n_\alpha  \alpha^{*} \Omega ^3 (\rho_{22}-\rho_{00})}{U_X^2\Gamma_x},\\ \label{g2rho1}
\rho_{12}\approx \frac{8 i \sqrt{2} n_\alpha  \alpha^{*}\tilde\Omega ^3 (\rho_{22}-\rho_{00})}{\tilde\Gamma_x}+24 \sqrt{2} n_\alpha\alpha^* \tilde\Omega ^3 (\rho_{11}-\rho _{00})\approx\frac{8 i \sqrt{2} n_\alpha  \alpha^{*}\Omega ^3 (\rho_{22}-\rho_{00})}{U_X^2\Gamma_x},\\ \label{g2rho2}
\rho_{02}\approx-\frac{4 i \sqrt{2} (\alpha^{*})^2 \tilde\Omega ^2 (\rho_{00}-\rho_{22})}{\tilde\Gamma_x}-4 \left(\sqrt{2} (\alpha^*)^2 \tilde\Omega ^2 (\rho_{00}-3\rho _{11}+2 \rho_{22})\right)\approx-\frac{4 i \sqrt{2} (\alpha^{*})^2 \Omega ^2 (\rho_{00}-\rho_{22})}{U_X\Gamma_x},
\end{gather}
where we have made the natural assumption that $\Omega\gg \Gamma_x$.

The equations for the diagonal elements of the density matrix are given by
\begin{gather}
\frac{d \rho_{00}}{dt}=-i \Omega  (\alpha \rho_{01}-\alpha^{*} \rho_{10})+\gamma_x \rho_{11}-\Gamma_x\rho_{00},\\
\frac{d \rho_{11}}{dt}=i \Omega  \left(\alpha(\rho_{01}-\sqrt{2}\rho_{12})+\alpha^{*}(- \rho_{10}+\sqrt{2} \rho_{21})\right)-\gamma_x  (\rho_{11}-2\rho_{22})+\Gamma_x (\rho_{00}-2\rho_{11}),\\
\frac{d\rho_{22}}{dt}=i \sqrt{2} \Omega  (\alpha \rho_{12}-\alpha^{*} \rho_{21})-2 \gamma_x \rho_{22}+2 \Gamma_x\rho_{11},
\end{gather}
we then replace the simplified expressions for the off-diagonal elements found in Eq.~\ref{g2rho}-\ref{g2rho2} to write the rate equations as follows 
\begin{gather}
\frac{d \rho_{00}}{dt}=+32\frac{ n^2_\alpha  \Omega ^4 (\rho_{22}-\rho_{00})}{U_X^2\Gamma_x}+\gamma_x \rho_{11}-\Gamma_x\rho_{00}+=8\frac{g_2^2}{\Gamma_x} n^2_\alpha (\rho_{22}-\rho_{00})+\gamma_x \rho_{11}-\Gamma_x\rho_{00},\\
\frac{d \rho_{11}}{dt}=-\gamma_x  (\rho_{11}-2\rho_{22})+\Gamma_x (\rho_{00}-2\rho_{11}),\\
\frac{d\rho_{22}}{dt}=-32\frac{ n^2_\alpha  \Omega ^4 (\rho_{22}-\rho_{00})}{U_X^2\Gamma_x}-2 \gamma_x \rho_{22}+2 \Gamma_x\rho_{11}=-8\frac{g_2^2}{\Gamma_x}n^2_\alpha  (\rho_{22}-\rho_{00})-2 \gamma_x \rho_{22}+2 \Gamma_x\rho_{11}.
\end{gather}
For $\gamma_x=0,$ the equation for $\rho_{11}$ simply gives $\rho_{00}=2\rho_{11}$ and yields the following equations
\begin{gather}
\frac{d \rho_{00}}{dt}=-\Gamma_x\rho_{00}+8\frac{g_2^2}{\Gamma_x} n^2_\alpha (\rho_{22}-\rho_{00}),\\
\frac{d\rho_{22}}{dt}= \Gamma_X\rho_{00}-8\frac{g_2^2}{\Gamma_x} n^2_\alpha (\rho_{22}-\rho_{00}),
\end{gather}
it is straightforward to note that  $\frac{d \rho_{00}}{dt}=-\frac{d \rho_{22}}{dt}.$ 

Finally, for the photon number we have the following equation
\begin{gather}
\frac{d n_\alpha}{dt}=-\gamma_c n_\alpha+i\Omega[\alpha^*(\rho_{10}+\sqrt{2}\rho_{21})-\alpha(\rho_{01}+\sqrt{2}\rho_{12})]\approx \\ \nonumber
\approx -\gamma_c n_\alpha+16\frac{g_2^2}{\Gamma_x} n^2_\alpha (\rho_{22}-\rho_{00}),
\end{gather}
from these equations  the two-photon emission rate is ready to be identified as   $A_{2}=8g_2^2/\Gamma_X.$ The equations can be written in its simplest form of
\begin{gather}
\label{ratee}
\frac{d \rho_{22}}{dt}\approx\Gamma_x\rho_{00}-A_2n_\alpha^2(\rho_{22}-\rho_{00}), \\ \nonumber
\frac{d n_\alpha}{dt}\approx-\gamma_cn_\alpha+2A_2 n_\alpha^2(\rho_{22}-\rho_{00}).
\end{gather}

The rate equations in Eq.~\ref{ratee} accept always a trivial solution with vanishing photon number $n_\alpha=0,$ and inverted population distributions $\rho_{00}=\rho_{11}=0$ and $\rho_{22}=1.$ However, beyond a critical two-photon emission rate a non-trivial steady state arises, this  critical $A_2^c$ is given by
\begin{gather}
A_2^{c}=\frac{5\gamma_c^2}{2\Gamma_x}, 
\end{gather}
and the photon number at the onset of two-photon gain is independent of $A_2$ 
\begin{gather}
n_\alpha=\frac{2\Gamma_x}{5\gamma_c}.
\end{gather}

One can further re-write the threshold for two-photon gain in terms of the two-photon coupling 
\begin{gather}
\frac{g^c_2}{\gamma_c}\approx \sqrt{\frac{5}{16}}\approx 0.55,
\end{gather}
where $g^c_2$ denotes the critical two-photon coupling.

In general, for finite $\gamma_x$, the trivial state has populations  given by  $\rho_{00}=\gamma_x^2/(\gamma_x^2+\gamma_x\Gamma_x+\Gamma_x^2)$ and $\rho_{11}=\gamma_x\Gamma_x/(\gamma_x^2+\gamma_x\Gamma_x+\Gamma_x^2).$  On the other hand, the appearance of the non-trivial state is slightly modified to
\begin{gather}
A_2^{c}=\frac{\gamma_c^2(5\Gamma_x+4\gamma_x)(\Gamma_x^2+\Gamma_x\gamma_x+\gamma_x^2)}{2(\Gamma_x^2-\gamma_x^2)^2},
\end{gather}
however, the photon number remains independent of $A_2$,
\begin{gather}
n_\alpha=2\frac{\Gamma_x^2-\gamma_x^2}{\gamma_c(5\Gamma_x+4\gamma_x)},
\end{gather}
note that solutions only appear for $\Gamma_X>\gamma_X$ as physically expected.

Finally, we shall remember that our numerical solutions work in the regime where $U_X\propto \Omega,$ that is, the exciton interactions  are of the order of magnitude of the light-matter coupling, leading to differences between the rate equations and the full numerical solutions. In addition, we have assume that the photon energy matches the cavity energy, $\omega=\omega_c$, contrary to the full numerics where $\omega$ is allowed to adapt self-consistently. However, the rate equations provide a valuable guide to understand the two-photon gain.
\end{widetext}
\end{document}